\documentclass[prx, letterpaper, superscriptaddress, amsmath, amssymb, amssymb, reprint, floatfix,onecolumn]{revtex4-2}

\usepackage{graphicx}
\usepackage{bm}
\usepackage{xcolor} 
\usepackage[normalem]{ulem}

\usepackage{braket}
\usepackage{esvect}

\begin{document}

\preprint{APS/123-QED}

\title{Electrical readout of spins in the absence of spin blockade}
\author{Felix-Ekkehard von Horstig}
\email{felix@quantummotion.tech}
\affiliation{Quantum Motion, 9 Sterling Way, London, N7 9HJ, United Kingdom}
\affiliation{Department of Materials Sciences and Metallurgy, University of Cambridge, Charles Babbage Rd, Cambridge CB3 0FS, United Kingdom}
\author{Lorenzo Peri}
\affiliation{Quantum Motion, 9 Sterling Way, London, N7 9HJ, United Kingdom}
\email{lp586@cam.ac.uk}
\affiliation{Cavendish Laboratory, University of Cambridge, JJ Thomson Ave, Cambridge CB3 0HE, United Kingdom}

\author{Virginia N. Ciriano-Tejel}
\affiliation{Quantum Motion, 9 Sterling Way, London, N7 9HJ, United Kingdom}

\author{Sylvain Barraud}
\affiliation{CEA, LETI, Minatec Campus, F-38054 Grenoble, France}
\author{Jason~A.~W.~Robinson}
\affiliation{Department of Materials Sciences and Metallurgy, University of Cambridge, Charles Babbage Rd, Cambridge CB3 0FS, United Kingdom}
\author{Monica Benito}
\affiliation{Institute of Physics, University of Augsburg, 86159 Augsburg, Germany}

\author{Frederico Martins}
\affiliation{Hitachi Cambridge Laboratory, J.J. Thomson Avenue, CB3 0HE, United Kingdom}
\author{M. Fernando Gonzalez-Zalba}
\email{fernando@quantummotion.tech}
\affiliation{Quantum Motion, 9 Sterling Way, London, N7 9HJ, United Kingdom}

\date{\today}

\begin{abstract}
In semiconductor nanostructures, spin blockade (SB) is the most scalable mechanism for electrical spin readout requiring only two bound spins for its implementation. In conjunction with charge sensing techniques, SB has led to high-fidelity readout of spins in semiconductor-based quantum processors. However, various mechanisms may lift SB, such as strong spin-orbit coupling (SOC) or low-lying excited states, hence posing challenges to perform spin readout at scale and with high fidelity in such systems. 
Here, we present a method, based on the dependence of the two-spin system polarizability on energy detuning, to perform spin state readout even when SB lifting mechanisms are dominant. It leverages SB lifting as a resource to detect selectively different spin measurement outcomes. We demonstrate the method using a hybrid system formed by a quantum dot (QD) and a Boron acceptor in a silicon p-type transistor and show spin selective readout of different spin states under SB lifting conditions due to (i) SOC and (ii) low-lying orbital states in the QD. We further use the method to determine the detuning-dependent spin relaxation time of 0.1-8~$\mu$s. Our method should help perform projective spin measurements with high spin-to-charge conversion fidelity in systems subject to strong SOC, will facilitate state leakage detection and enable complete readout of two-spin states. 

\end{abstract}

\maketitle

\section{Introduction}

\begin{figure}
    \centering
    \includegraphics[width=0.45\textwidth]{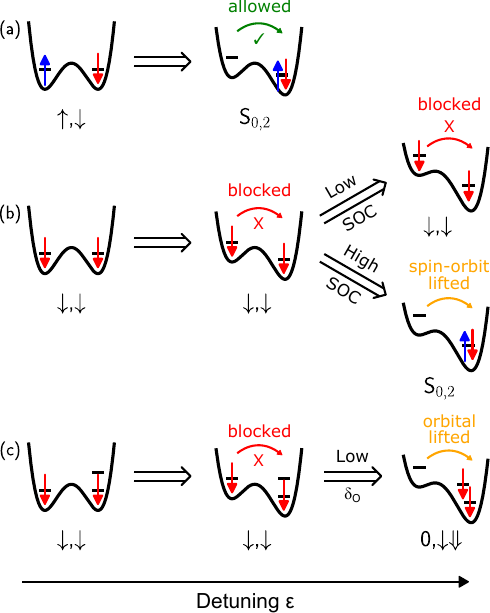}
    \caption{Spin readout using spin-blockade: (a) Anti-symmetric spins allow for the movement of spins, while (b,c) symmetric spins are blockaded, allowing for the two-particle spin states to be distinguished. At increased detuning, SB can be lifted in the presence of either large spin-orbit interaction (b) or the presence of low-lying excited orbital states of energy $\delta_o$ (c).}
    \label{fig:PSB-concept}
\end{figure}

Direct measurement of individual spins is an extremely challenging task given their small magnetic dipole moment. However, in semiconductor nanostructures, the charge dipole associated with electron tunnelling can be sizeable. Such divide in electronic properties is reflected in the vastly different state-of-the-art sensitivities for the spin ($\sim$ 10 spins/$\sqrt{\text{Hz}}$~\cite{Bienfait2016, Ranjan2020, Budoyo2020}) and charge degrees of freedom ($\sim$ 10$^{-6}$ electrons/$\sqrt{\text{Hz}}$~\cite{Schoelkopf1998, Aassime2001a, Schaal2020}) which has pushed researchers to develop spin-to-charge conversion (SCC) techniques in conjunction with charge sensing for electrical spin readout. 
Energy filtering, for example, uses the difference in tunnel rates to a charge reservoir of Zeeman-split spins states confined to a quantum dot (QD) (or impurity)~\cite{Elzerman2004} whereas spin blockade (SB) uses quantum selection rules to inhibit tunnelling between two-particle states with different spin numbers~\cite{Ono2002}, see Figure~\ref{fig:PSB-concept}. In particular, SB is a more scalable mechanism for SCC requiring only two bound spins for its implementation and has proven instrumental in achieving high fidelity readout of spin qubits~\cite{keith2019, Borjans2021, Blumoff2022, Oakes2023} even at low magnetic fields~\cite{Zhao2019} and high temperatures~\cite{Urdampilleta2019, Neigemann2022}. 

Charge sensing, in conjunction with SCC mechanisms as described above, relies on bi-state measurements arising from the charge sensor detecting either charge tunneling or the lack thereof. This means that only a specific subset of spin measurement outcomes may trigger the charge detector, while the absence of such trigger may be attributed to the complementary subset~\cite{Johnson2022}. However, this approach has drawbacks, including potential state leakage or spin mapping errors~\cite{Keith_2019b}. Furthermore, various mechanisms may lift SB, such as strong spin-orbit coupling (SOC) -- which allows two-particle states with different spin numbers to couple (Figure~\ref{fig:PSB-concept}b)~\cite{Danon2009,Nadj-Perge2010} -- and low valley-orbit level splitting -- allowing symmetric spin states to exist in different orbital or valley states of the same confining potential (Figure~\ref{fig:PSB-concept}c)~\cite{Shaji2008,Betz2015}. The former mechanism is particularly relevant to spins III-V materials~\cite{Wang2016, Maisi2016, Fujita2016, Wang2018} and novel spin qubit systems such as holes in germanium ~\cite{Watzinger2018, Jirovec2021,depalma2023strong, borsoi2024}, and silicon QDs~\cite{Li2015, Camenzind2022, Peri_vonHorstig_2024} that have come at the forefront of semiconductor based quantum computing due their technical ease for spin manipulation via electric fields~\cite{Maurand2016,Froning2021,Piot2022,schuff2024fully, carballido2024qubit} and their potential for large scale integration~\cite{Hendrickx2021,zhang2023universal}. A key element in the operation of spin qubits in double QDs is rapid spin projection(separation) pulses through a singlet-triplet anticrossing which are utilized to read(initialize) the system~\cite{liles2023singlettriplet}. However, the anisotropic nature of SOC on these systems results in magnetic field orientations in which SOC is strongly enhanced which may compromise satisfying the diabatic condition of the pulses and hence the ultimate achievable readout and initialization fidelity in scaled up systems (see Supplementary Note~1).

To overcome these challenges, we present a methodology, based on dispersive readout techniques~\cite{Vigneau2023} and spin-to-charge-polarizability conversion~\cite{West2019} that utilizes the very same coupling between states that leads to SB lifting, as a resource for selective readout of the states of a two-spin system. More concretely, our methodology makes use of the energy-detuning-dependent charge polarizability of the two-spin system to positively detect the spin measurement outcome without the need to perform a rapid diabatic pulse through a singlet-triplet anticrossing, minimising SCC mapping errors. The difference in polarizability manifests itself as a state-dependent quantum capacitance that can be detected through the dispersive interaction with a microwave superconducting resonator~\cite{vonhorstig2023}. We demonstrate this readout methodology using a hybrid system subject to strong SOC formed by a hole QD and a Boron acceptor in a silicon nanowire transistor and measure its spin relaxation time as a function of energy detuning. Finally, we expand the readout methodology to systems with low-lying orbital states, where SB may be lifted by allowed tunnelling to the higher energy orbital. We implement the demonstration in a different charge configuration of our hybrid system, and show that the spin states can be mapped onto signals arising from the orbital ground and excited state charge transitions, allowing for the spin states to be measured selectively. 

\section{Results}
\subsection{Spin blockade lifting via spin-orbit coupling}\label{sec:magneto-data}

\begin{figure*}[t]
    \centering
    \includegraphics[width=0.95\textwidth]{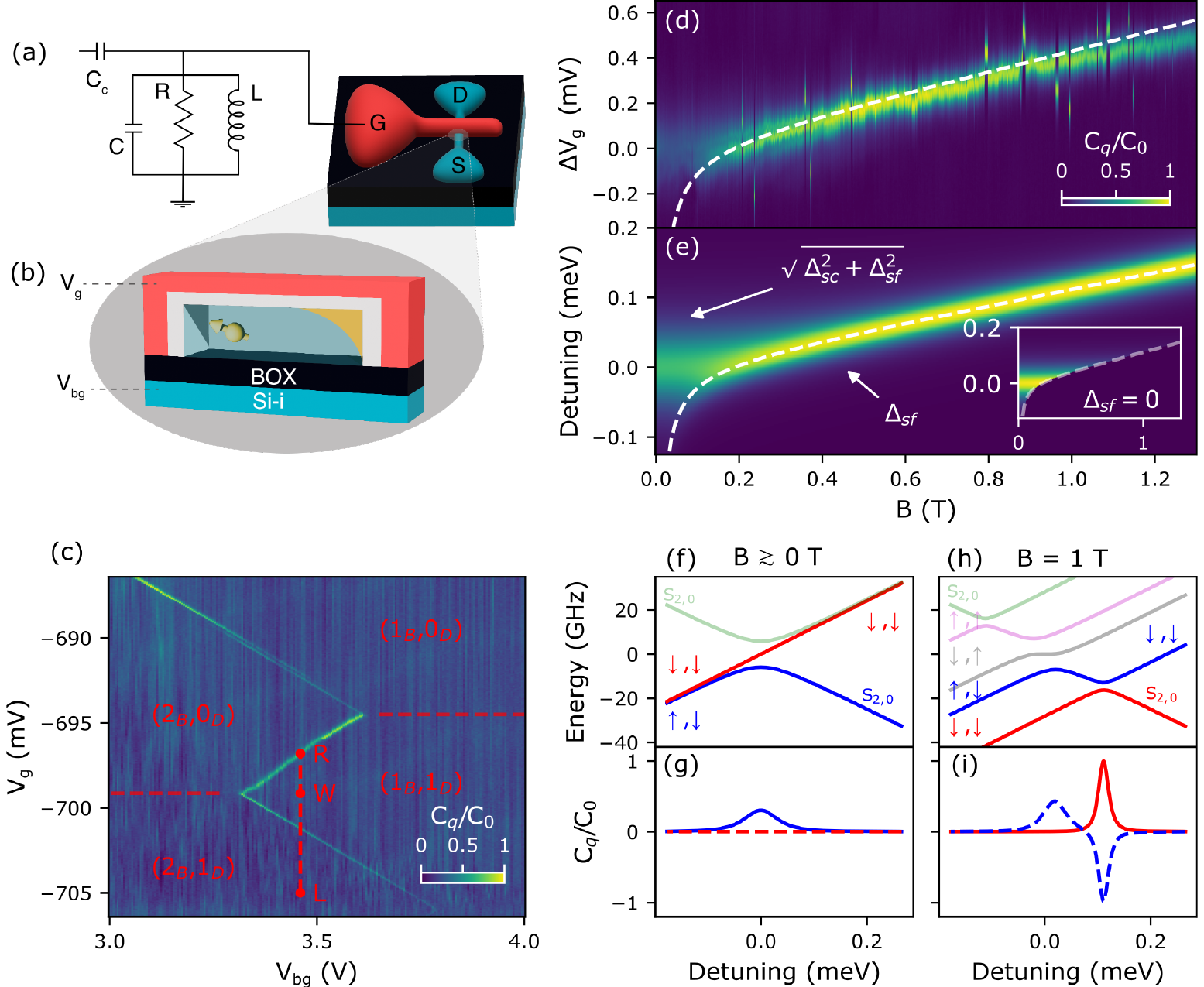}
    \caption{Device description and readout concept: (a)  
    Schematic of silicon nanowire transistor (top-view), labelled with Source (S), Drain (D) and top-gate (G) contacts, embedded in an LCR resonator for charge readout. (b) Schematic side-view of Si-nanowire with gate stack including gate metal (red), gate oxide (grey), channel (transparent blue), buried oxide (BOX in 
    black) and intrinsic silicon substrate (Si-i, also blue). The corner QD and Boron atom where holes are confined are marked in yellow. (c) Charge stability diagram showing the capacitive signal measured in the \textit{V}$_\text{g}$-\textit{V}$_\text{bg}$ space near the boron-dot transition (positive slope). A boron-reservoir transition is also visible (negative slope). The charge occupation are annotated in the plot. The approximate location of the dot-reservoir transition is indicated in red dashed lines. Location of Load, Wait and Read voltages for readout measurements are marked in red. (d) Magneto-spectroscopy measurement near the point labelled R in the stability diagram. The dotted line shows the location of the T$^-_{1,1}$/S$_{2,0}$ anticrossing.  (e) Simulated magneto-spectroscopy of the same transition. The insert shows a simulation of the same transition with $\Delta_\text{sf}$
    = 0, showing the emergence of SB. (f,h) Energy level diagram at two magnetic fields. 
    (g,i) Capacitive signal of the lowest two states ($\uparrow_B,\downarrow_D$) and ($\downarrow_B,\downarrow_D$) 
    at two magnetic fields. In each case the excited state is plotted in dashed lines.}
    \label{fig:readout-concept}
\end{figure*}

We use a p-type single-gate silicon transistor with light Boron channel doping, a system subject to strong SOC (Figure~\ref{fig:readout-concept}a)~\cite{Heijden2018}. By applying a voltage on the top-gate (\textit{V}$_\text{g}$) and back-gate (\textit{V}$_\text{bg}$), we accumulate holes in QDs formed in the nanowire, as well as in individual Boron atoms (Figure~\ref{fig:readout-concept}b). We connect the gate of the transistor to a superconducting microwave(mw) resonator (panel a), to detect quantum capacitance changes arising from mw-driven cyclic charge tunnelling between the QD/Boron and the source and drain charge reservoirs (S,D) as well as between QDs and Boron atoms. We refer to the latter as interdot charge transitions (ICTs).  

We tune the device to an ICT between a Boron atom and a QD (ICT A) with nominal charge occupation of ($N_\text{B}$,$N_\text{D}$) = (1,1)/(2,0) (Figure~\ref{fig:readout-concept}c) and a gate lever arm asymmetry $\Delta\alpha$~=~0.26~$\pm$~0.03, a parameter used to convert gate voltage to energy detuning between the QD and Boron atom, see Supplementar Note~2 for the full charge stability map. Throughout this work, we provide state occupations in the form (B,D), where the first will refer to the state of the Boron atom and the second to that of the QD. To characterise the spin eigenstates of the system, we perform magneto-spectroscopy~\cite{Lundberg2020} by measuring the resonator response, which at low temperature ($kB_T \ll \Delta_{sc}, \Delta_{sf}$ as defined below) corresponds to the quantum capacitance, $C_q$, of the ground state of the system, against gate voltage across the ICT and magnetic field strength applied in the plane of the sample (Figure~\ref{fig:readout-concept}d). 

The data reveals an enhancement of the resonator response which shifts up in $V_\text{g}$ as the magnetic field is increased above $B\approx0.2$~T. This is the signature of an effective two-spin system subject to strong SOC where, above 0.2~T, the system is free to tunnel between the polarised triplet state ($\downarrow_B,\downarrow_D$) and the joint singlet (S$_{2,0}$)~\cite{lundberg2021}. The enhancement in the signal originates from the reduced tunnel coupling of the $\downarrow_B,\downarrow_D$ (ground state above $B\approx0.2$~T) compared with the $\uparrow_B,\downarrow_D$ (ground state below $B\approx0.2$~T) with the S$_{2,0}$ state. This is SOC-mediated spin blockade lifting which we describe in detail below.

We consider the two-particle SOC Hamiltonian in Supplementary Note~3 that takes into account the $g$-factor of the Boron(QD) $g_\text{B(D)}$ (where in our case $g_\text{B} < g_\text{D}$) and the spin-conserving and spin-flip tunnel coupling, $\Delta_\text{sc}$ and $\Delta_\text{sf}$. 
In Figure~\ref{fig:readout-concept}e, we plot the simulated magneto-spectroscopy, as well as the eigenenergies and capacitive signals (Figure~\ref{fig:readout-concept} f,h and g,i) arising from the lowest two states at $B = 0(1)$~T. For $B<0.2$~T, the capacitive signal arises from charge tunneling between the ground ($\uparrow_B,\downarrow_D$) 
and S$_{2,0}$ states, mediated by the finite $\Delta_\text{sc}$, see blue trace in panel g. For $B>0.2$~T, however, the signal arises from the tunneling between the Zeeman split  T$^-_{1,1}$ = ($\downarrow_B,\downarrow_D$) and the S$_{2,0}$ state, in this case mediated by $\Delta_\text{sf}$
, see the red trace in panel i. The good match between the experiment and simulations confirms the presence of the lifting mechanism since, for systems with low SOC, like few-electron QDs in silicon, such spin-flip tunneling processes are forbidden resulting in the signal vanishing asymmetrically at high fields (see the simulation in the insert of Figure~\ref{fig:readout-concept}e)~\cite{Mizuta2017}.

Magneto-spectroscopy allows us to quantify the parameters in the Hamiltonian. First, we obtain the average $g$-factor of the hole spin in the QD and Boron, $\overline{g} \sim$ 2, which is given by the slope of the transition in the T$_{1,1}^-$/S$_{2,0}$ regime ($\overline{g}=e\alpha\Delta V_\text{g}/\mu_B B$), where $e$ is the electron charge and $\mu_B$ the Bohr magneton~\footnote{We define the $g$-factor at a particular magnetic field angle as $g = |\boldsymbol{g}\vec{B}|/|\vec{B}|$ where $\boldsymbol{g}$ is the $g$-tensor.}. Further, from the linewidth of the signal at zero and high field (T$_{1,1}^-$/S$_{2,0}$ regime), we extract the total charge and spin-flip coupling rates, $\sqrt{2} \Delta_c = $  $ \sqrt{\Delta_\text{sc}^2 + \Delta_\text{sf}^2} \sim$ 10.4~GHz and $\Delta_\text{sf} \sim$ 3.4~GHz, respectively. Additionally, by measuring the shift in $V_\text{g}$ of the Boron-reservoir transition with magnetic field (not shown), we find the $g$-factor of the Boron $g_\text{B} \sim$ 1.6 making the $g$-factor of the QD, $g_\text{D} \sim$ 2.4.

\subsection{Spin readout under spin-orbit lifted SB}\label{sec:ST-readout-demo}

\begin{figure}
    \centering
    \includegraphics[width=0.45\textwidth]{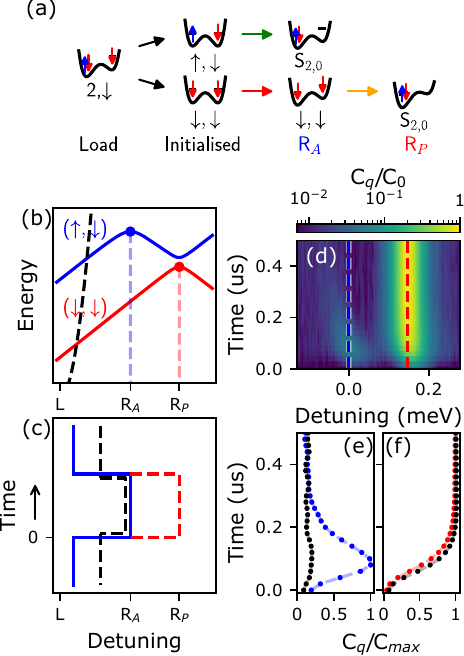}
    \caption{Readout protocol with spin-orbit coupling: (a) Schematic of pulsing scheme. The state is initialised by randomly unloading a spin from the Boron atom starting from a (2,$\downarrow$) configuration resulting in either a ($\downarrow_B,\downarrow_D$), or a ($\uparrow_B,\downarrow_D$) The measurement is performed either at R$_A$ [triggered by the ($\uparrow_B,\downarrow_D$) state] or at R$_P$ [triggered by the ($\downarrow_B,\downarrow_D$) state].  (b) Energy level diagram showing the load (L) and read points for the two states (R$_A$, R$_P$). The black dashed line denotes the Boron-reservoir transition between the (2,1) and the (1,1) charge states. (c) Schematic of pulse sequence for spin readout. States are initialised at L, and then pulsed to either R$_A$ or R$_P$ (blue or red lines). The black dashed lines indicates the control pulse where the state is initialised in the (1,1) region. The time at which data acquisition is started is indicated. (d) Capacitive signal as a function of measurement time and detuning showing two signal arising from the ($\downarrow_B,\downarrow_D$) and ($\uparrow_B,\downarrow_D$) states. R$_A$ and R$_P$ are indicated with dashed lines. Measurements carried out at $B = 1$~T. (e,f) Line-cuts of d at detuning of R$_A$ and R$_P$ showing the capacitive signal (normalised to the peak value of each trace) against time. Dashed lines are given as a guide to the eye. The black lines are the signals recorded from the control measurement. }
    \label{fig:readout-proof-of-principle}
\end{figure}

Spin blockade is based on the exclusion principle in which two fermionic particles cannot posses all the same quantum numbers. It is utilized in DQDs to project the combined spin state of two separated spin carrying particles into that of a single QD whose ground state is the joint singlet, $S_{2,0}$. Particularly, the polarized Triplets $T^\pm_{1,1}$ are blockaded from transitioning into $S_{2,0}$ a feature that is used for spin parity readout, or for Singlet-Triplet readout if the unpolarized triplet state $T^0$ is also blocked~\cite{Seedhouse2021}. 

In the presence of SOC, however, $\Delta_\text{sf}$ enables tunnelling of the $T^\pm_{1,1}$ states into the $S_{2,0}$, eliminating the spin selectivity of SB readout via projective measurements unless a fast diabatic pulse across the $\Delta_\text{sf}$ anticrossing can be performed. Such a requirement may be challenging in practice given that the $\Delta_\text{sc}$ anticrossing also needs to be crossed adiabatically, ultimately limiting the fidelity of the SCC mechanism (see Supplementary Note~1).

In this work, we use $\Delta_\text{sf}$ to our advantage. The finite spin flip coupling term between the T$^-_{1,1}$ and S$_{2,0}$ spin branches generates a distinct anticrossing at positive energy detuning $\varepsilon$. Most critically, for the purpose of our demonstration, we note that the capacitive signals arising from the anticrossings of the aligned ($\downarrow_B,\downarrow_D$) and anti-aligned ($\uparrow_B,\downarrow_D$) spin outcomes occur at different detuning (see the red and blue traces in Figure~\ref{fig:readout-concept}i respectively), a concept that we exploit in the following for spin readout. 

We note two additional features of this readout scheme that may be used to distinguish the other spin states, T$^+_{1,1}$ = ($\uparrow_B,\uparrow_D$) and ($\downarrow_B,\uparrow_D$), from the T$^-_{1,1}$ and ($\uparrow_B,\downarrow_D$) states: First, the T$^+_{1,1}$ and ($\downarrow_B,\uparrow_D$) additionally anticross with the  S$_{0,2}$ state at a third and fourth readout point at negative detuning (see Figure~\ref{fig:readout-concept}h), allowing for these populations to be distinctly measured (Supplementary Note~4). Second, beyond a positive and negative $C_q$, also a neutral response can be measured. This allows for the spin states producing a $C_q$ to be additionally distinguished from those that do not (see Supplementary Note~5). For example, assuming the spin states were adiabatically transferred from the (1,1), at the anticrossing between the T$^-_{1,1}$ and S$_{2,0}$ states, a positive (or negative) $C_q$ would correspond to a T$^-_{1,1}$ (or ($\uparrow_B,\downarrow_D$)) measurement outcome respectively, while a neutral response indicates either a T$^+_{1,1}$ or ($\downarrow_B,\uparrow_D$) state. Together, these two techniques may be used to measure the full spin state of the two-spin system (Supplementary Note~4.1).

To measure the spin of the Boron, we first randomly initialise the system in either ($\uparrow_B,\downarrow_D$) or T$^-_{1,1}$. We do so by starting in the (2$_B$,1$_D$) charge state (point L in Figure~\ref{fig:readout-concept}c) to then pulse into the (1$_B$,1$_D$) region (point W), which randomly unloads a spin from the Boron atom, as illustrated in Figure~\ref{fig:readout-proof-of-principle}a. We then pulse to either of the readout points (R$_A$, R$_P$ where the subscripts stand for anti-parallel and parallel) and measure the dispersive signal in the time domain, see Figure~\ref{fig:readout-proof-of-principle}d-f. For these preliminary experiments, we set the wait time at W to $t_\text{W}=0$~s. Additionally, we perform a control measurement in which we wait in the (1,1) region to deterministically initialize the system in the T$^-_{1,1}$ by relaxation (see pulse sequences in Figure~\ref{fig:readout-proof-of-principle}c and Methods). 

We observe two signals at different detuning (panel d), the separation of which depends on the magnetic field intensity as anticipated  (see Supplementary Note~6)
. When we take a cut at zero detuning (panel e), we observe the signal from the ($\uparrow_B,\downarrow_D$) - S$_{2,0}$ anticrossing, i.e. a ($\uparrow_B,\downarrow_D$) measurement outcome. The signal initially rises, due to the finite ring up of the resonator, before decaying with a time constant $T_1 \sim$ 100~ns given by the relaxation time to the T$^-$ state (blue trace). We highlight a slower decay in the signal from the ($\uparrow_B,\downarrow_D$) measurement at more negative detuning, which we attribute to a detuning-dependent T$_1$ which we investigate further in the following Section. We note that the short $T_1$ at the readout point prevented us from performing single-shot readout measurements (see Supplementary Note~7 for further discussion).

At finite positive detuning (R$_P$), on the other hand, we observe signals arising from the T$^-_{1,1}$-S$_{2,0}$ anticrossing, i.e. a T$^-_{1,1}$ measurement outcome (Figure~\ref{fig:readout-proof-of-principle}f). In this case, the signal (red trace) is delayed with respect to the resonator ring up (black trace). The slower dynamics is caused by the fraction of ($\uparrow_B,\downarrow_D$) shots that carry a negative quantum capacitance at R$_P$, hence reducing the signal at timescales comparable to the relaxation time of 95~ns in this case. 

Our result shows that, in spin systems with lifted SB due to SOC, the spin state can be read using the different detuning points at which the dispersive signal of the ($\uparrow_B,\downarrow_D$) and T$^-_{1,1}$ measurement outcomes manifest. This measurement is done without the need to perform a perfectly diabatic pulse through the T$^-_{1,1}$-S$_{2,0}$ anticrossing.

\subsection{Spin relaxation time} 

\begin{figure}
    \centering
    \includegraphics[width=0.45\textwidth]{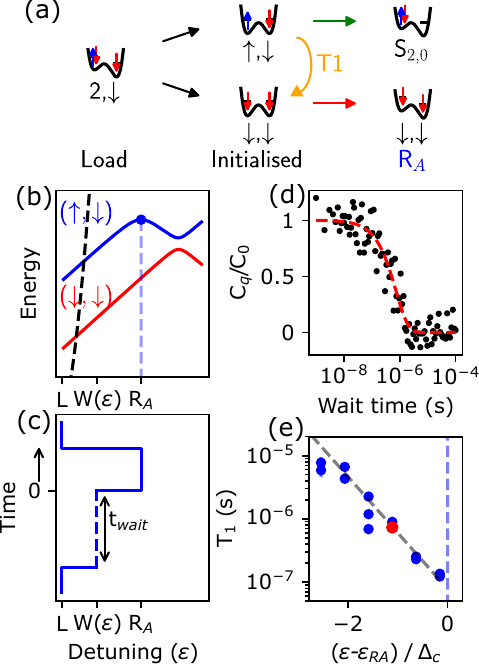}
    \caption{$T_1$ characterisation: Measurements carried out at $B = 700$~mT. (a-c) Schematic of the measurement protocol: The state is initialised as described in Figure~\ref{fig:readout-proof-of-principle}. This is followed by a Wait (W($\varepsilon$)) period of variable time and pulse depth in which ($\uparrow_B,\downarrow_D$) may relax into T$^-_{1,1}$. Finally the state is read out at the spin anti-parallel readout point (R$_A$). The location of W($\varepsilon$) is varied to characterise $T_1$ as a function of $\varepsilon$. The time at which data acquisition is started is indicated. (d) Example capactive signal [a proxy for the ($\uparrow_B,\downarrow_D$) population] against t$_{\text{w}}$ at $\varepsilon$ = -92 $\mu$eV = -1.12 $\Delta_c$ (red dot panel e). The data is taken from the maximum signal of line traces similar to that in Figure~\ref{fig:readout-proof-of-principle}d. The data is fitted using an exponential decay to extract T$_1$ (730~$\pm$~80~ns in this case). (e) $T_1$ against detuning of the wait location W($\varepsilon$) showing an exponential dependence (black dashed line) with detuning. The location of R$_A$ is marked in blue. The data in panel d is marked with a red dot.}
    \label{fig:T1-vs-detuning}
\end{figure}

To demonstrate the benefit of this spin readout mechanism, we now study the spin qubit decay constant $T_1$ as a function of detuning. We again initialise randomly in the ($\uparrow_B,\downarrow_D$) or T$^-_{1,1}$ state but wait in the (1,1) region (point $W(\varepsilon)$) for a variable time before pulsing to the readout point R$_A$ (Figure~\ref{fig:T1-vs-detuning}a-c). For long $t_\text{W}$, the initialised state will have a higher chance to decay to the ground state resulting in a reduction in the average excited state signal (Figure~\ref{fig:T1-vs-detuning}d). We extract $T_1$ by fitting the data to an exponential decay of the form $P_S(t_\text{W}) \propto \exp(-t_\text{W}/T_1)$ (red dashed line). We repeat this measurement for different detuning points and find that $T_1$ increases exponentially away from the readout point (Figure~\ref{fig:T1-vs-detuning}e) up to 8 $\mu$s, increasing its utility as a spin qubit. 

We hypothesize the short $T_1$ at the readout point is caused by the coupling of the spin and charge degrees of freedom, i.e. at the anticrossing, the spin-carrying charges have the largest dipole and therefore are most impacted by charge noise. We note that a charge relaxation time, $T_1$ = 100~ns has been measured in a similar sample \cite{Urdampilleta2015}.

To further support the hypothesis, we discard spin-orbit coupling as the dominant source of relaxation at the readout point due to lack of dependence of $T_1$ with magnetic field \cite{Burkard2023_Semispinsreview} (see Supplementary  Note~6), where we show several time-dependent readout signal similar to that reported in Figure~\ref{fig:readout-proof-of-principle} but recorded at different magnetic fields.

\subsection{Spin readout under orbitally lifted SB}
\label{sec:Excited_State}
\begin{figure*}
    \centering
    \includegraphics{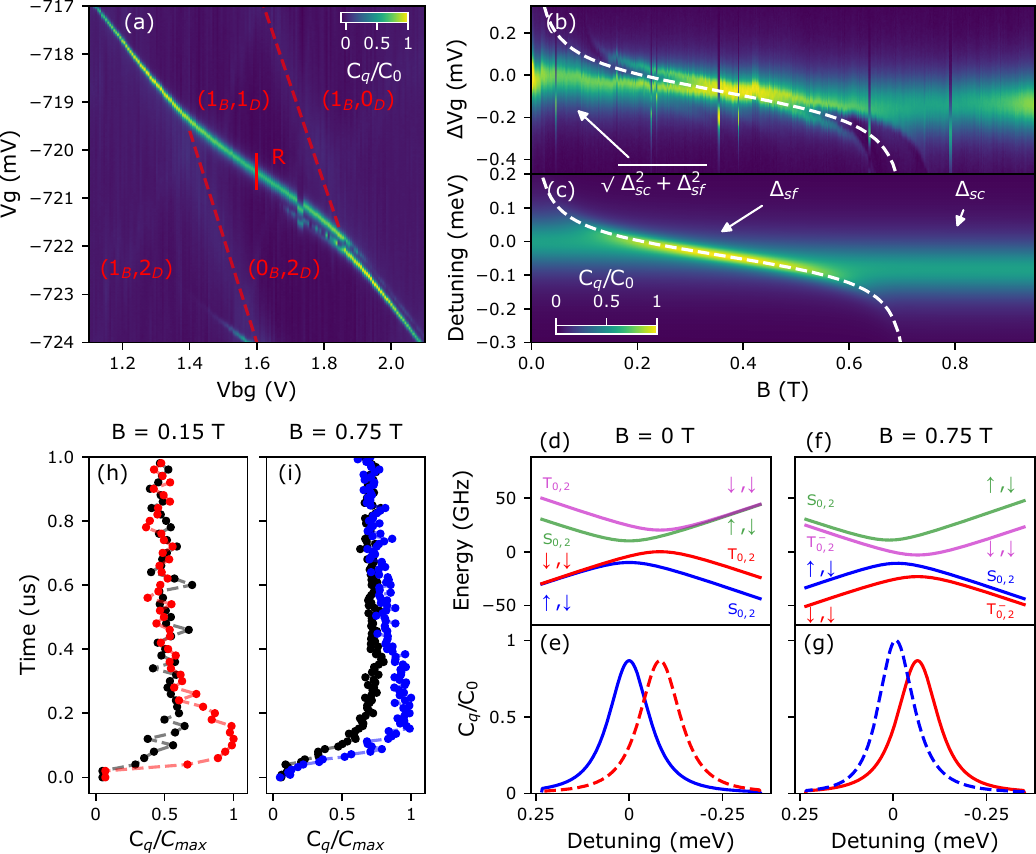}
    \caption{Readout utilising orbital states: (a) Stability diagram of ICT B showing the nominal charge occupation. The readout location is highlighted. (b,c) Magneto-spectroscopy and simulation at the point marked R in a. (d,f) Energy level diagrams showing the energy levels of the transition at two magnetic fields. (e,g) Capacitive signals from the ($\uparrow_B,\downarrow_D$)/S$_{0,2}$ (blue) and T$^-_{1,1}$/T$^-_{0,2}$ (red) anticrossings marking the two readout locations. In each case the excited state is marked in dashed lines. (h,i) Capacitive signal measured, used to distinguish ($\uparrow_B,\downarrow_D$) and T$^-_{1,1}$ states utilising the orbital T$^-_{1,1}$/T$^-_{0,2}$ transition (low magnetic field) or the ($\uparrow_B,\downarrow_D$)/S$_{0,2}$ transition (high magnetic field) as readout points. The response is normalised to the maximum of each linetrace.
    }
    \label{fig:orbital-readout}
\end{figure*}

In projective SB measurements, the presence of low-lying excited orbital or valley states (of energy $\delta_o$) lifts SB by allowing spin triplets (T$_{0,2}$) to exist in the (0,2) charge configuration. Hence, when the energy detuning exceeds $\delta_o$, both triplet and singlet are allowed to transition into the (0,2), eliminating spin selectivity of the charge movement (Figure~\ref{fig:PSB-concept}c). Although techniques are available to increase the energy of the $\delta_o$, for example by moving to a (1,3) charge occupation to fill the lowest valley in electron systems \cite{Harvey-collard2017} or by reducing the size of the QD to increase orbital energies, the excited states still impact the readout of using projective SB. This mechanism limits the magnetic fields at which SB can be performed in transport measurements to $B<\delta_o/(\overline{g}\mu_\text{B})$~\cite{Shaji2008}, while in charge-sensing experiments it limits the size of the voltage window in which SB can be detected ~\cite{Weinstein2023}.

In this Section, we demonstrate that the readout mechanism described in the context of SOC, naturally extends to anticrossings arising from orbital states. In particular, we show that instead of being a constraint, the presence of orbital states can be used as a resource by resulting in two distinct readout locations which can be used to measure the spin state of the hybrid DQD. Such approach allows to selectively and positively detect different spin measurement outcomes. While this method does not increase the size of the readout window as compared to standard charge-sensed SB, it provides an alternate methodology to read the spin state dispersively.

For the demonstration, we tune the device to a different ICT comprising a different Boron atom and QD (ICT B) with nominal charge occupation ($N_\text{B}$,$N_\text{D}$) = (1,1)/(0,2) and differential lever arm $\Delta\alpha$ = 0.54 $\pm$~0.02 (Figure~\ref{fig:orbital-readout}a). Notably, in the (0,2) configuration the two holes reside within the QD, in contrast to ICT A where they resided in the Boron. The presence of low-energy orbital excited states in the QD allow for an orbital lifting of SB. We perform magneto-spectroscopy and again find an enhancement of the signal at magnetic fields above 300~mT characteristic of a two-spin-system with SOC. However, in this case, we additionally find that above 600~mT the signal remains at a fixed $V_\text{g}$ point and its intensity is reduced, the signature that a transition from T$^-_{1,1}$ to the T$^{-}_{0,2}$ state involving an orbital excitation in the QD is now allowed~\cite{Betz2015}. 

From magnetospectroscopy (Figure~\ref{fig:orbital-readout}b), we extract $\Delta_c\sim$~20~GHz, $\Delta_\text{sf}\sim$~4~GHz, $\delta_o\sim$~20~GHz, where $\delta_o$ is the QD excited state energy. From the slope of the transition in the T$_{1,1}^-$/S$_{0,2}$ and the T$^{-}_{0,2}$ regimes, we extract the average $g$-factor, $\overline{g} = 2.1 \pm 0.1$,  and $g$-factor difference $\delta g = g_B - g_D = 0.3 \pm 0.1$, respectively.
We extend the Hamiltonian to include the $T^-_{0,2}$ state (Supplementary Note~3) and simulate the magnetospectroscopy in Figure~\ref{fig:orbital-readout}c. In the data, we note the additional edges in the signal parallel to the T$_{1,1}^-$/S$_{0,2}$ anticrossing (white dashed line), see Figure~\ref{fig:orbital-readout}b. These arise due to resonant interactions between the spin system and photons in the mw resonator (2.1~GHz). Although less clear, these can also be observed in Figure~\ref{fig:readout-concept}d.

We plot the energy-level diagrams and capacitance from the ground and first excited states at $B = 0$~T and $B = 0.75$~T  in Figure~\ref{fig:orbital-readout}d-g. For ease of readability, we only include the T$_{1,1}^-$ and ($\uparrow_B,\downarrow_D$) states in the (1,1) charge region and the T$_{0,2}^-$ in the (0,2) region. We further set $\Delta_\text{sf}=0$, since it is less relevant at fields explored in this section, i.e. outside the intermediate field regime where the T$^-$/S anticrossing occurs. At magnetic fields below and above the T$^-$/S regime, two distinct readout points emerge corresponding to the ($\uparrow_B,\downarrow_D$)/S$_{0,2}$ anticrossing (ground state at low magnetic field), and the T$^-_{1,1}$/T$^{-}_{0,2}$ anticrossing (ground state at high magnetic fields), separated in detuning by $\delta_o + ([g_B - g_D^*]/2)\mu_\text{B}B$, where $g_D^*$ is the $g$-factor of the excited state of the doubly occupied QD. 

We now demonstrate readout in each of these regimes. First, at low field ($B=0.15$~T), we use the T$^-_{1,1}$/T$^-_{0,2}$ anticrossing as the readout point. We start deep in the (1,1) region, where the ground state is T$^-_{1,1}$, to then perform a diabatic passage through the T$_{1,1}^-$/S$_{0,2}$ anticrossing to prepare the system in the excited T$^-_{1,1}$ state near zero detuning. Finally, we ramp to the T$^-_{1,1}$/T$^-_{0,2}$ anticrossing and gather the time-domain response (red points in Figure~\ref{fig:orbital-readout}h). We observe an initial resonator ring-up followed by a decay with $T_1$ $\sim$ 140~$\pm$~50~ns (see Methods). In this case, the signal does not fully decay to zero as it was the case for the SOC experiments. This is a particularity of our concrete experiments since the signal from the ground state can also be detected at the readout point since $\delta_o \lesssim \Delta_\text{c}$ (note the overlap in the C$_q$ peaks in Figure~\ref{fig:orbital-readout}e and g). We discuss this limitation further in Supplementary  Note~9. We compare the signal to a control measurement where we initialise in the ground state deep in the (0,2) region by waiting for relaxation to the S$_{0,2}$ state. We then ramp to the readout point (black points). In this case, the system remains in the ground state and the signal rises with the ring up of the resonator.  

For the high field case (B=0.75~T), where we use the ($\uparrow_B,\downarrow_D$)/S$_{0,2}$ anticrossing for readout, we perform a similar sequence but starting deep in the (0,2) region where the ground states is S$_{0,2}$. Via diabatic pulsing through the T$^-$/S$_{0,2}$ anticrossing (outside the detuning range of panel f), we prepare the system in the excited S$_{0,2}$ near zero detuning to then ramp to the readout point. We observe a similar resonator ring-up followed by a decay with now a $T_1$ $\sim$ 270~$\pm$~50~ns. We plot the data and the control in Figure~\ref{fig:orbital-readout}i. 

By measuring the decay constant as a function of detuning near the readout points, we find that $T_1$ ranges between 100-200~ns (200-400~ns) in the low (high) magnetic field case, with larger values reported further from the readout point. We hypothesize that the difference in $T_1$ for the two cases is related to the state decay happening primarily between the T$_{0,2}$ and S$_{0,2}$ at low fields - a spin decay within the QD - while at high magnetic fields the decay occurs between the ($\uparrow_B,\downarrow_D$) and ($\downarrow_B,\downarrow_D$) states - a decay within the Boron atom.

Overall, our results show that, in spin systems with lifted SB due to orbital states, the spin state can be read selectively and positively by making use of the different detuning points at which the dispersive signal of the ($\uparrow_B,\downarrow_D$) and T$^-_{1,1}$ measurement outcomes manifest. While SOC is present in our system, resulting in a finite $\Delta_\text{sf}$ and $g$-factor difference, neither are required for readout using orbital states, allowing the readout mechanism to be extended to systems lacking strong SOC such as electrons in silicon. In such systems ($\Delta_\text{sf}$ = 0, $g_D$ = $g_B$), the two readout points correspond to a Singlet or Triplet (T$^\pm$, T$^0$) outcome. We discuss this further in Supplementary Note~4.2.

\section{Discussion}\label{sec:discussion}

We have presented a novel spin readout methodology based on the detuning-dependent polarizability of the two-spin system in a semiconductor DQD to perform spin readout even when SB lifting mechanisms are present. We demonstrate this readout mechanism in two situations: Readout in the presence of SOC leading to spin flip tunnel coupling between the T$^-$ and S$_{2,0}$ states, and spin-blockade lifting due to the presence of excited orbital states.

We note that, going beyond the applications presented in the experimental section, our methodology enables three-valued readout, where up to three different spin sub-states can be discerned with a single measurement at a given measurement point (see Supplementary Note~5). This is a consequence of the charge polarization nature of the method where a positive, negative and neutral response of the system can be detected. The tri-state and selective nature of of our measurements may therefore enable complete readout of the state of two-spin systems more efficiently than what is possible through charge sensing~\cite{Nurizzo2023}, reducing the number of necessary repetitive measurements~\cite{Philips2022} (see Supplementary Note~4.1).

Our work and methodology opens new opportunities to (i) study the fundamentals of SB, its angular dependence in SOC systems and its impact on the ultimate readout fidelity. We highlight the limit of typical charge-sensed SB protocols to perform SCC due to incomplete diabatic passage of the $\uparrow,\uparrow$/S$_{2,0}$ anticrossing (see Supplementary Note~1). Since the method presented in this work does not require any diabatic passages it may be a way to enable high SCC fidelity. (ii) The ability to use the selective and tri-state nature of the readout may be used to enhance the spin readout fidelity, detect state leakage and complete readout of two-spin systems without the need of perturbative tunnel barrier pulses~\cite{Nurizzo2023}. In particular, by successively measuring at the various readout points of the spin-orbit coupled system, the full spin state, as well as state leakage out of the computational subspace may be detected (see Supplementary  Note~4.1). (iii) Our work encourages the exploration of hybrid QD-acceptor systems and its interaction with a microwave resonator as a system for quantum information processing. This work shows interesting promise that in these systems the spin-relaxation time may be improved, increasing their utility as potential qubits. The acceptor (which can be seen as a hole analogue to a phosphorus donor) produces tight confinement, increasing the energy of orbital states.

\section*{Methods}
\textbf{Fabrication details.} The transistors used in this study consists of a single gate silicon-on-insulator (SOI) nanowire transistor with a channel width of 120~nm, a length of 60~nm and height of 8~nm on top of a 145-nm-thick buried oxide. The silicon layer has a Boron doping density of $5\cdot 10^{17}$~cm$^{-3}$ . The silicon layer was patterned to create the channel using optical lithography, followed by a resist trimming process. The transistor gate stack consists of 1.9~nm HfSiON -- leading to a total equivalent oxide thickness of 1.3~nm -- capped by 5~nm TiN and 50~nm polycrystalline silicon. After gate etching, a Si$_3$N$_4$ layer (10 nm) was deposited and etched to form a first spacer on the sidewalls of the gate, then 18-nm-thick Si raised source and drain contacts were selectively grown before source/drain extension implantation and activation annealing. A second spacer was formed, followed by source/drain implantations, an activation spike anneal and salicidation (NiPtSi). The nanowire transistor and superconducting resonator were connected via Al/Si 1\% bond wires.

\textbf{Measurement set-up.} Measurements were performed at the base temperature of a dilution refrigerator ($T\sim$~10~mK). Low frequency signals ($V_\text{g}$, $V_\text{bg}$) were applied through Constantan twisted pairs and $RC$ filtered at the MXC plate. Radio-frequency signals were applied through filtered and attenuated coaxial lines to a coupling capacitor at the input of the $LC$ resonator. Fast pulsing signals were applied through attenuated coaxial CuNi lines to an on-PCB (printed circuit-board) bias-T connected to the source of the transistor. The resonator (characteristic frequency 2.1~GHz) consists of a NbTiN superconducting spiral inductor (L $\sim$ 30~nH), coupling capacitor ($C_c \sim$ 40~fF) and low-pass filter fabricated by \textit{Star Cryoelectronics}. For exact details of the superconducting chiplet see ref. \cite{vonhorstig2023}. The PCB was made from 0.8-mm-thick RO4003C with an immersion silver finish. The reflected rf signal was amplified at 4~K and room temperature, followed by quadrature demodulation (Polyphase Microwave AD0540B), from which the amplitude and phase of the reflected signal were obtained (homodyne detection).

\textbf{Readout pulse sequence.} The data in Figure~\ref{fig:readout-proof-of-principle}d consists of 100,000 shots at each detuning point, resulting in an average resonator response of the initialised states. After the readout measurement cycle in Figure~\ref{fig:readout-proof-of-principle}c, we wait for 100 microseconds in the (1,1) region to ensure the spin in the QD has decayed to the $\downarrow$ ground state. This ensures only the ($\uparrow_B,\downarrow_D$) and ($\downarrow_B,\downarrow_D$) states can be initialised. For the time domain data (Figure~\ref{fig:readout-proof-of-principle}d-f), the time constant of the resonator ring-up ($\tau \sim$ 80~ns) is in good agreement with the bandwidth of our resonator ($\kappa/2\pi$ $\sim$ 3-4~MHz with the exact value depending on magnetic field, and $\tau$ = 2/$\kappa$). For readout at R$_A$, we estimate the state decay constant $T_1$ from the exponential decay of the signal after the initial rise. For readout at R$_P$, the initialised signal rises more slowly ($\tau \sim$ 95~ns) as compared to the control ($\tau \sim$ 80~ns). 

\textbf{Fit of time traces to extract $T_1$ at the readout point.}
To extract the state decay constants from the time-traces in Figure~\ref{fig:readout-proof-of-principle}~e and Figure~\ref{fig:orbital-readout}h and i, we fit the data with a model combining the ring-up of the resonator (determined by $\tau \sim$ 80~ns) with capacitive signal contributions arising from the ground ($C_\text{gnd}$) and excited state ($C_\text{exc}$) where the excited state exponentially decays into the ground determined by a time constant $T_1$:

\begin{equation}
    C_\text{total} = \left(1-e^{\frac{-t}{\tau}}\right) \left[C_\text{exc}e^\frac{-t}{T_1} +  C_\text{gnd} \left(1 - e^\frac{-t}{T_1}  \right) \right],
\end{equation}

\noindent where $t$ is the measurement time. We use a least square fit to extract $T_1$ and estimate the uncertainty from the covariance of the fit.
\par


\section*{Acknowledgement}
This research was supported by the UK's Engineering and Physical Sciences Research Council (EPSRC) via the Cambridge NanoDTC (EP/L015978/1). F.E.v.H. acknowledges funding from the Gates Cambridge fellowship (Grant No. OPP1144). J.W.A.R. acknowledges funding from the EPSRC Core-to-Core International Network Grant “Oxide Superspin” (No. EP/ P026311/1). M.F.G.Z. acknowledges a UKRI Future Leaders Fellowship [MR/V023284/1]. L.P. acknowledges support from The Winton Programme for the Physics of Sustainability. MB acknowledges funding from the  Emmy Noether Programme of the German Research Foundation (DFG) under grant no. BE 7683/1-1.

\section*{Author contributions}
F.E.v.H. acquired and analysed the data under the supervision of M.F.G.Z. and J.W.A.R. and F.M.. F.E.v.H., L.P. and M.F.G.Z. conceived and designed the experiment and contributed to the writing of the manuscript. L.P. developed the spin modelling under supervision by M.F.G.Z and M.B.. S.B. fabricated the CMOS device.

\section*{Data availability}
The source data that support the plots within this article and other findings of this study are provided with this study. 

\section*{Competing interests}
The authors declare no competing interests.

\bibliography{General_clean}


\end{document}


\preprint{APS/123-QED}

\title{Supplementary - Electrical readout of spins in the absence of spin blockade}
\author{Felix-Ekkehard von Horstig}
\email{felix@quantummotion.tech}
\affiliation{Quantum Motion, 9 Sterling Way, London, N7 9HJ, United Kingdom}
\affiliation{Department of Materials Sciences and Metallurgy, University of Cambridge, Charles Babbage Rd, Cambridge CB3 0FS, United Kingdom}
\author{Lorenzo Peri}
\affiliation{Quantum Motion, 9 Sterling Way, London, N7 9HJ, United Kingdom}
\email{lp586@cam.ac.uk}
\affiliation{Cavendish Laboratory, University of Cambridge, JJ Thomson Ave, Cambridge CB3 0HE, United Kingdom}

\author{Virginia N. Ciriano-Tejel}
\affiliation{Quantum Motion, 9 Sterling Way, London, N7 9HJ, United Kingdom}
\author{Sylvain Barraud}
\affiliation{CEA, LETI, Minatec Campus, F-38054 Grenoble, France}
\author{Jason~A.~W.~Robinson}
\affiliation{Department of Materials Sciences and Metallurgy, University of Cambridge, Charles Babbage Rd, Cambridge CB3 0FS, United Kingdom}
\author{Monica Benito}
\affiliation{Institute of Physics, University of Augsburg, 86159 Augsburg, Germany}

\author{Frederico Martins}
\affiliation{Hitachi Cambridge Laboratory, J.J. Thomson Avenue, CB3 0HE, United Kingdom}
\author{M. Fernando Gonzalez-Zalba}
\email{fernando@quantummotion.tech}
\affiliation{Quantum Motion, 9 Sterling Way, London, N7 9HJ, United Kingdom}

\date{\today}

\maketitle

\appendix

\renewcommand{\figurename}{}
\renewcommand{\thefigure}{Supplementary Figure \arabic{figure}}
\renewcommand{\thesection}{Note \arabic{section}}
\renewcommand{\theequation}{\arabic{section}}
\renewcommand{\thesubsection}{\arabic{section}.\arabic{subsection}}

\section{Spin-to-charge conversion fidelity in charge-sensed SB}\label{app:sec:disc-convet-PSB-limits}

\begin{figure}[tb!]
    \centering
    \includegraphics[width=\linewidth]{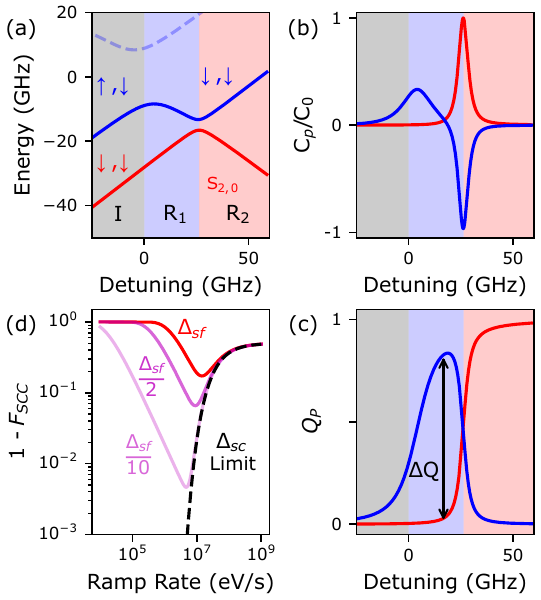}
    \caption{Comparison of charge-sensed SB and dispersive sensing. (a) Energy-level diagram of the system in Section \ref{sec:magneto-data}, showing the initialisation and operation regime (grey), and two potential readout regimes for SB measured via charge sensing (blue and red areas). (b-c)  Capacitance $C_p$ (i.e. polarisability) and charge polarisation $Q_p$ plotted against detuning for the parameters plotted in (a). (d) Limitations in $F_\text{SCC}$ based on incomplete adiabatic transitions when measuring in readout regime $R_2$, calculated for the parameters in (a) plotted in solid red and for those with $\Delta_\text{sf}$ reduced by a factor of 2 and 10 in the shaded lines. }
    \label{fig:psb-comparison}
\end{figure}

In charge-sensed SB, the spin system is ramped from within the (1,1) (grey area marked I in \ref{fig:psb-comparison}a) into the (2,0) charge configuration (red area marked $R_2$). In systems with negligible SOC (i.e. $\Delta_\text{sf}$, $\delta g$ = 0), this process is typically performed using a single slow, adiabatic passage (SAP) across the charge anticrossing ( i.e. $S_{1,1}/S_{2,0}$) which converts the singlet population into the (2,0) charge configuration, while keeping the triplets in the (1,1) \cite{Taylor2007}. In the presence of $\Delta_\text{sf}$, additionally, a fast ramp across the $T^-_{1,1}$/$S_{2,0}$ anticrossing may diabatically transition the system into the typical states (i.e. map the $T^-_{1,1}$ to a (1,1) and the ($\uparrow,\downarrow$) into the (2,0) charge configuration). Such ramp must be sufficiently large to achieve a diabatic transition with high probability and hence a high spin-to-charge conversion fidelity, F$_\text{SCC}$. It is worth highlighting that any loss of fidelity caused by this state projection correspond to errors in the SCC mechanism, which limit the fidelity independently of the sensor's SNR and system's $T_1$.

Physical limitations in the maximum achievable ramp rate may place an upper limit on the diabatic transfer fidelity. Based on a 1~V pulse over 1~ns attenuated by 30 dB, with a $\Delta\alpha$ of 0.26, we estimate that the maximum achievable pulse rate in our system is 8.2~MeV/s, limiting the fidelity to diabatically cross the $\uparrow,\downarrow$/$S_{2,0}$ anticrossing to below 79\%.  However, even if a faster ramp rate (compared to $\Delta_\text{sf}$) is achievable, it must also be kept low enough so as not to diabatically excite the states across the $\uparrow,\downarrow$/$S_{2,0}$ anticrossing, which incorrectly maps the ($\uparrow,\downarrow$) into the (1,1) charge configuration, therefore reducing F$_\text{SCC}$.

For the charge-sensed mode of operation, F$_\text{SCC}$ is then simply the probability of neither process producing an incorrect spin-to-charge mapping \cite{Peri_vonHorstig_2024}, which for a constant ramp rate $\nu$ is given by:

\begin{equation}\label{eq:app:fscc-psb}
    F_{SCC} = e^{-2\pi\frac{\Delta_\text{sf}^2}{\hbar\nu}} \left(1 - \frac{1}{2}e^{-4\pi\frac{\Delta_\text{sc}^2}{\hbar\nu}} \right),
\end{equation}

\noindent Note that by modulating $\nu$ it may be possible to achieve an improved $F_{SCC}$ but nevertheless the figure calculated from Eq.~\ref{eq:app:fscc-psb} can be used as an indication of the limitation imposed by needing to adiabatically transition the spin conserving and diabatically transitioning the spin-flip transition. Additionally, for large $\Delta_\text{sf}/\Delta_\text{sc}$, finding an optimal ramp sequence may not be trivial at every magnetic field given the potential anticrosing overlap of the $\Delta_\text{sf}$ and $\Delta_\text{sc}$ anticrossings.

For the parameters of $\Delta_\text{sf}$ and $\Delta_\text{sc}$ described in Section~\ref{sec:magneto-data}, we plot the infidelity as a function of $\nu$ in \ref{fig:psb-comparison}b (dark red trace). At low $\nu$, the fidelity is limited by the probability to cross the $\Delta_\text{sf}$ anticrossing, while at high $\nu$ the fidelity is limited by the $\Delta_\text{sc}$ anticrossing (black-dashed line). For the experimental parameters, a maximum fidelity of 68~\% can be achieved at the optimal ramp rate of 7~MeV/s. A reduction in $\Delta_\text{sf}$, which may be achieved by tuning the magnetic field angle \cite{Yu2023}, can increase the fidelity as demonstrated by the additional, shaded magenta lines in \ref{fig:psb-comparison}d for a reduction of $\Delta_\text{sf}$ by a factor of 2 and 10 respectively. It is worth noting that any $\Delta_\text{sf}$, regardless of how small, will result in a finite infidelity 1-F$_\text{SCC}$, limiting the maximum achievable fidelity by this method of diabatically ramping across the $\downarrow,\downarrow$/$S_{2,0}$ anticrossing using a constant ramp rate.

Two additional modes of operation may be envisioned to get around this issue: (i) performing the measure before the $\downarrow,\downarrow$/$S_{2,0}$ anticrossing, or (ii) attempting to traverse the entire energy landscape adiabatically. 

\begin{enumerate}

\item By measuring between the two anticrossings (blue area marked $R_1$ in \ref{fig:psb-comparison}a), the problem of imperfect diabatic transitions can be avoided as only the ($\uparrow,\downarrow$)/$S_{2,0}$ transition needs to be crossed adiabatically. However, if the anticrossings are not sufficiently separated, readout fidelity may be limited by their overlap. In \ref{fig:psb-comparison}c, we plot the charge polarisation $Q_p$ of the two lowest states, corresponding to the average excess charge on the second dot, as a function of detuning. Note that due to the overlap of the anticrossings, the charge polarisation falls short of unity for the $\uparrow,\downarrow$ state, resulting in a reduced contrast to the T$^-_{1,1}$. This overlap also affects dispersively sensed SB (upper panel) but the impact is less severe. We discuss the impact of this overlap further in Supplementary~\ref{app:sec:readout-visibility-separation}.

\item Another mode of operation may be envisioned where both anticrossings are crossed adiabatically and a charge sensor measurement is taken in the region marked $R_2$ in \ref{fig:psb-comparison}a. In this case, the $T^-_{1,1}$ is mapped into the $S_{2,0}$ state (i.e. a (2,0) charge configuration), while the $\uparrow,\downarrow$ is mapped into the $T^-_{1,1}$ state (i.e. a (1,1) charge configuration) -- all other states, $T^+_{1,1}$ and $\downarrow,\uparrow$ are also mapped into the (1,1) charge configurations. This charge mapping is opposite to the usual SB. In this case, the state-projection fidelity is limited by state decay occurring during the time it takes to perform the (slow) adiabatic ramp.

\end{enumerate}

\section{Stability map showing transitions A and B} \label{app:sec:stability-map}
\begin{figure}
    \centering
    \includegraphics[width=0.5\textwidth]{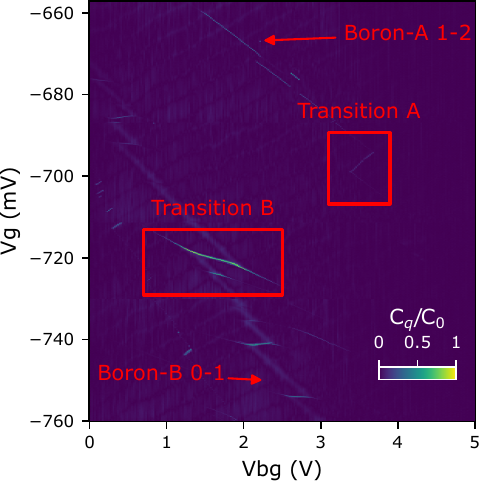}
    \caption{Stability map showing the relative location of ICT A and B.}
    \label{fig:app:stability-map}
\end{figure}

The stability map of the device under study shows two Boron-reservoir transitions (red arrows) intersected by several Boron-quantum-dot-transitions. The different Boron acceptors and QDs can be distinguished by their different slope with respect to \textit{V}$_\text{bg}$ and \textit{V}$_\text{g}$ indicating different gate lever arms $\alpha$. The charge occupation of the Boron atoms was deduced from the magneto-spectroscopy. Since Boron atoms can hold at most two holes, and the Boron involved in ICT A already contains two, we conclude that the Boron atoms involved in ICT A and B are distinct Boron atoms. This is supported by the difference in slope (different $\alpha$) and tunnel rate of the two Boron-reservoir transitions.

\section{Modelling Energy levels and Magneto-spectroscopy} \label{app:sec:Model}
We introduce the following Hamiltonian to simulate the energy level diagrams and magneto-spectroscopy in this work. We consider first two charge islands with occupation ($N_L$,$N_R$) occupied by a total of two spins. 

In the following, we assume the left dot is singly/not occupied, while the right dot is singly/doubly occupied. In the presence of SOC, it is easier to describe the (1,1) occupation in the single spins basis ($|\uparrow_l,\uparrow_r\rangle$, $|\downarrow_l,\uparrow_r\rangle$, $|\uparrow_l,\downarrow_r\rangle$, and $|\downarrow_l,\downarrow_r\rangle$). In the (0,2) occupation, we only include the singlet $|S_{0,2}\rangle$~\cite{Veldhorst2015}. In this basis, the Hamiltonian of the system can be found in Eq.~(\ref{eq:H_SO_5X5}), which is a generalization of Ref.~\cite{Veldhorst2015,Danon2009}. The Zeeman energies are determined by the two g factors $g_l$ and $g_r$ for the two QDs respectively, while the (1,1) region is connected to the $|S_{0,2}\rangle$ via a spin-conserving ($\Delta_\text{sc}$) and a spin-flip ($\Delta_\text{sf}$) tunnel couplings \cite{Yu2023}, the latter arising from the presence of SOC and thus potentially different spin quantization axes between the left and right QDs \cite{sen2023,vanriggelen-doelman2023}.
\begin{widetext}
    \begin{equation}
          H =
         \frac{1}{2}
        \begin{bmatrix}
                \varepsilon + (g_l + g_r) B & 0 & 0 & 0 & -\Delta_\text{sf}\\
                0 &  \varepsilon + (- g_l + g_r) B & 0 & 0 & -\Delta_\text{sc}\\
                0 & 0 & \varepsilon + (g_l - g_r) B & 0 & \Delta_\text{sc}\\
                0 & 0 & 0 & \varepsilon + (-g_l - g_r) B & \Delta_\text{sf}\\
                -\Delta_\text{sf} & -\Delta_\text{sc} & \Delta_\text{sc} & \Delta_\text{sf} & - \varepsilon\\
        \end{bmatrix}
        \label{eq:H_SO_5X5}
    \end{equation}
\end{widetext}

This Hamiltonian allows us to calculate the energy level diagrams shown in this paper. To simulate the magneto-spectroscopy, we extract the capacitive signal of the ground state arising from the cyclical variations in charge occupations driven by the tone of the resonator. This gives rise to a parametric capacitance (quantum and tunneling capacitance~\cite{Mizuta2017}). We neglect contributions by relaxation events giving rise to Sisyphus resistance~\cite{Esterli2019}. The signal arises whenever two states anticross as this allows the resonator to drive differences in the charge occupation of the QDs. For a two level system with tunnel coupling $\Delta_0$ in the slow relaxation limit (negligible tunneling capacitance), this results in a capacitance of:

\begin{equation}\label{eq:app:Cq}
    C_p = \frac{(e\alpha)^2}{2} \frac{\Delta_0^2}{\left[(\varepsilon - \varepsilon_0)^2 + \Delta_0^2\right]^{3/2}}\chi_c
\end{equation}
\noindent

where $C_p$ is the parametric capacitance, $\alpha$ the lever arm,  $\varepsilon_0$ the location of the anticrossing in detuning, and $\chi_c$ the charge polarisation. It is worth noting that a smaller $\Delta_0$ will result in sharper and brighter transitions (as long as $\Delta_0 > f_r$ \cite{Esterli2019,peri2024a}). In the limits described, 
an intuitive understanding of the capacititve signal arising from a given energy level diagram can be gained from the second derivative (i.e. the curvature) of the energy level with detuning arising from the anticrossing of two states~\cite{Petersson2010}. 

At zero magnetic field, the spin anti-parallel to S$_{0,2}$ anticrossing dominates the ground state of the system and therefore provides the source of the capacitive signal. Once a magnetic field is applied, the T$^-_{1,1}$ state is lowered below the ($\uparrow,\downarrow$) and ($\downarrow,\uparrow$) states which for $g_r-g_l \ll \overline{g}$ and $\Delta_{sf} \ll \Delta_{sc}$ can be found analytically as $\varepsilon \gtrsim \left[\Delta_{sc}^2 - ( 2\overline{g}\mu_B B)^2\right]/(4\overline{g}\mu_B B)$, otherwise the solution can be found numerically. Once $2\overline{g}\mu_B B \gtrsim \Delta_{sc}$ this results in the T$^-_{1,1}$ cutting off the anti-parallel to S$_{0,2}$ signal at zero detuning. If $\Delta_{sf} > 0$, the T$^-_{1,1}$ will then anticross with the S$_{0,2}$ giving rise to a new capacitive signal whose centre follows the hyperbolic expression above. 

In the presence of an orbital (or valley) excited state in the right dot (Section~\ref{sec:Excited_State}), separated from the ground state in the right by an energy $\delta_o$, we can extend the model in Eq.~(\ref{eq:H_SO_5X5}) to include the state $|T^-_{0,2}\rangle$. Once $2g_r  
\mu_B B \gtrsim \delta_o$,
the T$^-_{0,2}$ is also lowered below the S$_{0,2}$ state. At this point, the ground state anticrossing switches from the T$_{1,1}^-$/S$_{0,2}$ to the T$^-_{1,1}$/T$^-_{0,2}$ at $\varepsilon$ = $\delta_o$ + $1/2$~$(g_l - g_r)\mu_B B$. 
For the sake of simplicity of the model, we assume that the overlap of the two states with the left QD are similar enough not to meaningfully alter the tunnel coupling, or the $g$-factor. Therefore, we assume $\Delta_\text{sc}$ and $\Delta_\text{sf}$ to be the same for both $|S_{0,2}\rangle$ and $|T^-_{0,2}\rangle$, as confirmed by the experimental measurement in Section~\ref{sec:Excited_State}.


\section{Readout locations for hole and electron systems}\label{app:sec:readout-locations}

In this section, we will discuss how different anticrossings can be used for state readout and the impact of magnetic field on the anticrossing separation of these states. For the purposes of clarity, we will split this section into two 
regimes: (i) Systems with strong SOC (e.g. hole spins in Si, Ge), resulting in significant $\Delta_\text{sf}$ but where we neglect low-lying orbital states, and (ii) systems with weak SOC (e.g. electrons spins in Si), where $\Delta_\text{sf}$ = 0 and $g_l = g_r$ but low-lying orbital or valley states are taken into account. Holes in QDs can exhibit both SOC and low lying orbital states (not discussed here) leading to a combination of the two effects.

\subsection{Readout in systems with strong SOC and full state readout}\label{app:sec:SOC-readout-full}
\begin{figure*}[htb]
    \centering
    \includegraphics[width=1\textwidth]{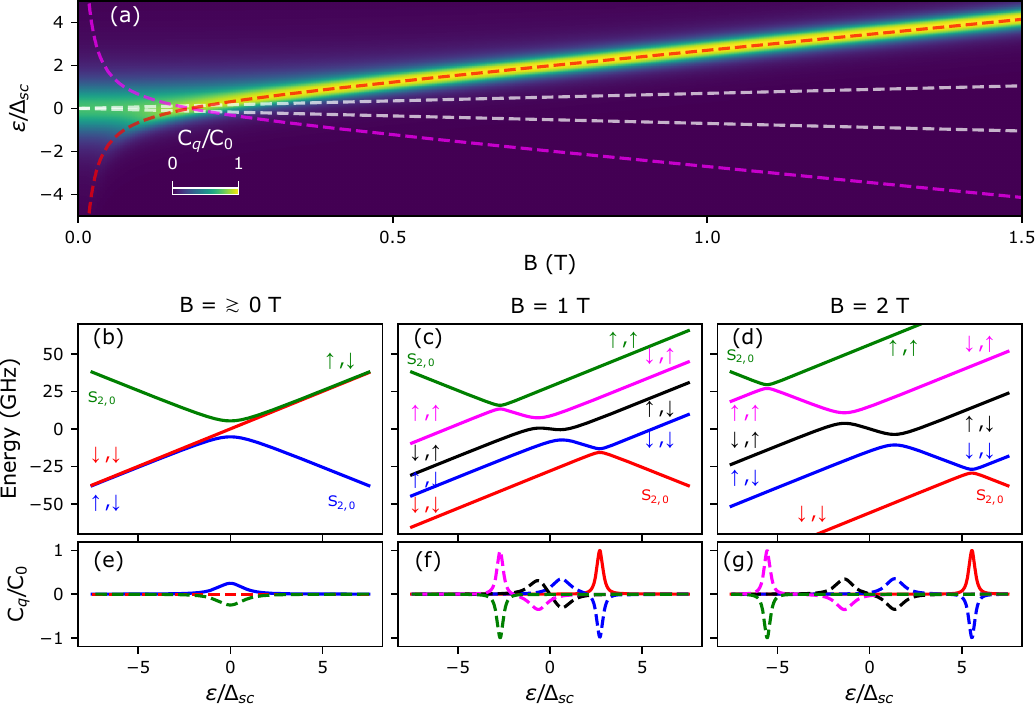}
    \caption{Readout locations for systems with strong SOC: a) Simulated magneto-spectroscopy of a two spin system with $\Delta_\text{sf}$ = 10 GHz, $\Delta_\text{sf}$ = $\Delta_\text{sf}$/4, and $g_l$ = 1.5 ,$g_r$ = 2.5. The location of the T$_{1,1}^-$/S$_{2,0}$, T$_{1,1}^+$/S$_{2,0}$ anticrossings are marked with dashed lines in red and pink while those for the ($\uparrow,\downarrow$)/S$_{2,0}$ and ($\downarrow,\uparrow$)/S$_{2,0}$ are marked in white. b,c,d) Energy level diagrams at B = 0, 1~T and 2~T with d,e,f) the quantum capacitance arising from each state states. At elevated magnetic field there are four readout points corresponding to the T$^-_{1,1}$, ($\uparrow,\downarrow$), ($\downarrow,\uparrow$) and T$^+_{1,1}$ states.}
    \label{fig:app:hole-acceptors-energies}
\end{figure*}

To discuss the readout points available for systems with strong SOC, we plot a simulated magneto-spectroscopy map, energy level diagrams, and quantum capacitance signals for a DQD system with $\Delta_\text{sc} = 10$~GHz, $\Delta_\text{sf} = \Delta_\text{sc}/4$, and $g_l$ = 1.5, $g_r$ = 2.5 (Fig. \ref{fig:app:hole-acceptors-energies}). In this case, we do not consider orbital excited states.

At elevated magnetic fields, the T$^-_{1,1}$, ($\uparrow,\downarrow$), ($\downarrow,\uparrow$) and T$^+_{1,1}$ each anticross with the S$_{2,0}$ (\ref{fig:app:hole-acceptors-energies}d and g), giving rise to four distinct readout points which can be used to perform full state readout. Given a long enough $T_1$, combined with slow enough ramp rates to ensure adiabatic passage of the anticrossings ($\hslash\nu \ll \Delta$), the distinct readout points could allow a full characterization of the state of the two spin system in a complementary way to what has been demonstrated in Ref.~\cite{Nurizzo2023}. 

The readout sequence for full two-spin state might involve successively adiabatically ramping to all or a subset of the readout points occurring at the anticrossings between each of the spin states and the S$_{2,0}$, and measure the quantum capacitance, $C_q$, via a resonator response. Note that some of the states will produce $C_q$ responses at multiple readout points: for example the T$^+_{1,1}$ state will produce a positive $C_q$ at the T$^+_{1,1}$ to S$_{0,2}$ anticrossing, but will produce a negative $C_q$ at the ($\downarrow,\uparrow$) to S$_{0,2}$ anticrossing. Since the sign of the anticrossing differs however, these cases can be distinguished as they produce an opposite dispersive shift on the resonator. 

In the minimal case, a measurement at the ($\downarrow,\uparrow$) to S$_{0,2}$ anticrossing allows the T$^+_{1,1}$ state (negative $C_q$) and ($\downarrow,\uparrow$) state (positive $C_q$) to be distinguished from each other and from the remaining ($\uparrow,\downarrow$) and T$^-_{1,1}$ states (no $C_q$). This could be followed by a measurement at the T$^-_{1,1}$ to S$_{0,2}$ anticrossing which produces a negative $C_q$ for the $\uparrow,\downarrow$, a positive $C_q$ for the T$^-_{1,1}$, and no $C_q$ for the remaining ($\downarrow,\uparrow$) and T$^+_{1,1}$ state. Together, these two measurements are able to distinguish the four states from each other. 

The ability to subsequently measure the occupation of the states may be used to detect state leakage out of the computational subspace. If between the individual measurement steps, the spin state decays or state leakage out of the computational subspace occurs (e.g. into higher-orbital states or due to the loss or gain of an additional charge), this error can also be detected by an erroneous response of the measurement protocol. For example, a positive $C_q$ at the first measurement step (indicating a ($\downarrow,\uparrow$) state) followed by a positive $C_q$ in the second measurement step (indicating a T$^-_{1,1}$ state) could indicate a decay in the second spin. More thorough measurement procedures may be able to detect more or different kinds of errors at the cost of longer measurement time. For example, negative $C_q$ recorded at the T$^+_{1,1}$ to S$_{0,2}$ anticrossing would indicate the spin state had leaked out of the computational subspace into S$_{0,2}$.

To allow for sufficient visibility between the states, the readout locations need to be separated from each other to avoid signal overlap, as was discussed in Supplementary~\ref{app:sec:readout-visibility-separation}. This can be done either by changing the tunnel coupling, as discussed before, or by increasing the magnetic field: The T$^+$/S and T$^-$/S each move in detuning with magnetic field described by $\varepsilon_{\pm} = \pm \left[\Delta_c^2 - ( 2\overline{g}\mu_B B)^2\right]/4\overline{g}\mu_B B$ (see red and pink dashed lines in Fig. \ref{fig:app:hole-acceptors-energies}a). For the anti-aligned states, the separation is determined by the difference in g-factor $\delta g = g_r - g_l$ giving $\varepsilon_{(\uparrow,\downarrow)/(\downarrow,\uparrow)}$ $\approx$ $\pm \delta g \mu_B B/2$ (white dashed lines).  

In order to allow for sufficient fidelity in the adiabatic passage, the ramp rate to transition onto and between the readout locations needs to be sufficiently slow as determined by the Landau-Zener equation. While in principle the time to perform the slow ramps may allow for state decay to occur, the large charge coupling typical of dispersive readout significantly reduces this time. If, for example a detuning range of 100~GHz needs to be crossed with a anticrossing of 1~GHz in the range, a ramp rate of 17.7~keV/s (taking 2.3~$\mu$s) could be used to achieve an infidelity of only $10^{-4}$.

\subsection{Readout in systems with orbital excited states} \label{app:sec:orbital-readout-full}

\begin{figure*}
    \centering
    \includegraphics[width=1\textwidth]{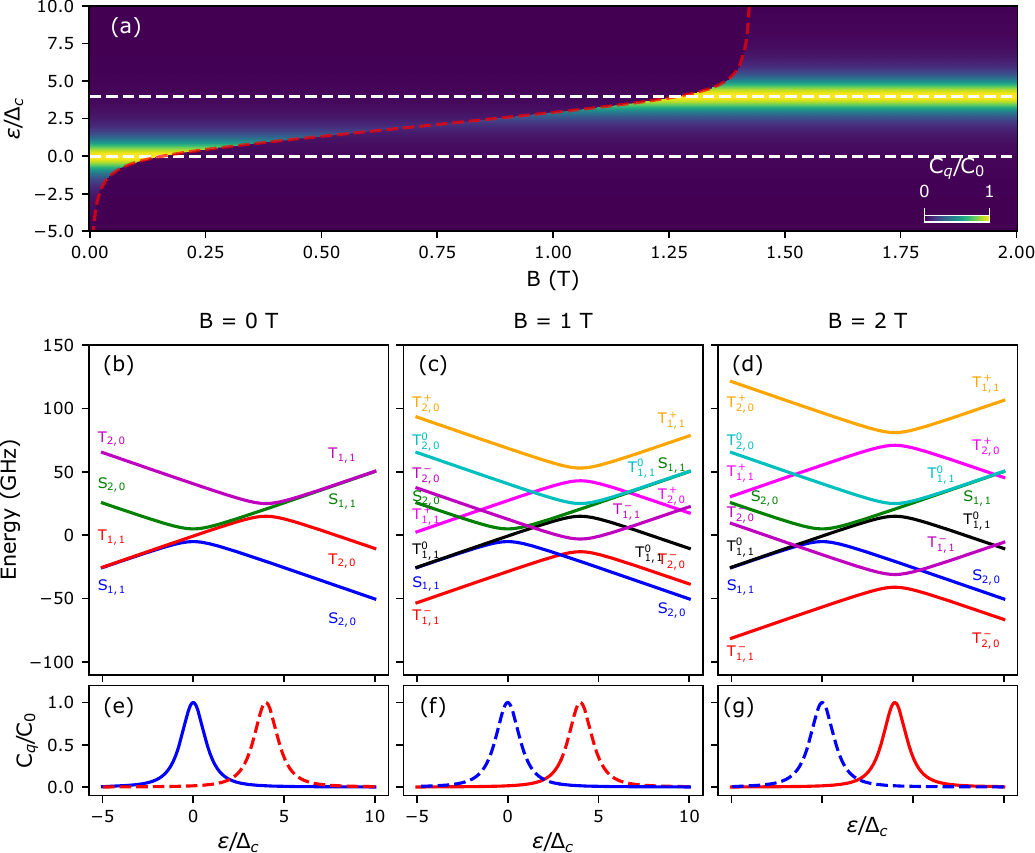}
    \caption{Readout locations for systems with orbital excited states: a) Simulated magneto-spectroscopy of a electron double quantum dot with $\Delta_c$ = 10 GHz, $\delta_o$ = 4$\Delta_c$ and $g_l,g_r$ = 2 ($\Delta_\text{sf}$ = 0). The location of the S/S (left), and T/T (right) anticrossings are marked with white dashed lines, the location of the T$^-$/S crossing is marked with a red dashed line. b,c,d) Energy level diagrams at B = 0, 1T and 2T with e,f,g) the quantum capacitance arising of the S$_{1,1}$ and T$_{1,1}$ states. In each case the capacitance arising from the instantaneous ground state is plotted in a solid line, while excited states are plotted in dashed lines. There are two readout points corresponding to the S$_{1,1}$ and T$_{1,1}$ outcomes.}
    \label{fig:app:electrons-energies}
\end{figure*}

Low-lying orbital (or valley) states are common in gate-defined QDs and their effect on the readout positions should be considered. While, it was more appropriate to describe the case of holes in Si in the single spin basis due to their sizable g-factor difference, for electron in Si, where g-factors tend to be isotropic and SOC is weak, it is more practical to use the Singlet-Triplet basis. We define the singlet and unpolarized triplet as S$_{1,1}=(\downarrow,\uparrow-\uparrow,\downarrow)/\sqrt{2}$ and T$^0_{1,1}=(\downarrow,\uparrow + \uparrow,\downarrow)/\sqrt{2}$, respectively. To provide a complete discussion, we additionally include the three Triplet states in the (0,2) charge occupancy involving the orbital excited state.

We plot a simulated magneto-spectroscopy map, energy level diagrams and quantum capacitance signals for a DQD housing two spins with $\Delta_c = 10$~GHz, $\delta_o = 4\Delta_c$ and $g_l,g_r = 2$ ($\Delta_\text{sf} = 0$) in Fig. \ref{fig:app:electrons-energies}. The existence of the orbitally excited T$_{0,2}$ states significantly increases the complexity of the energy level diagrams (Fig. \ref{fig:app:hole-acceptors-energies}c,d,e). However, the quantum capacitance only arises from a few anticrossings: S$_{1,1}$/S$_{0,2}$, T$^-_{1,1}$/T$^-_{0,2}$, T$^0_{1,1}$/T$^0_{0,2}$, T$^+_{1,1}$/T$^+_{0,2}$, which are marked in blue/green, red/magenta, black/cyan and pink/orange respectively. In principle the three T/T anticrossings shift in magnetic field according to their g-factor difference ($g_l$ - $g_r$) but since we have set all g-factors to be the same (as is typical for electron in Si) this does not occur here.  As a result, only two distinct readout locations are present arising from the S/S anticrossing at $\varepsilon$ = 0 and from the three T/T anticrossings located at $\varepsilon$ = $\delta_o$ (Fig. \ref{fig:app:electrons-energies}c,d,e). These readout points are present at all magnetic fields, allowing for singlet-triplet readout independent of the magnetic field intensity. 

To increase readout visibility, the separation of the readout locations should be maximised. These are fixed with magnetic field in this case. We therefore require the orbital (or valley) energy $\delta_o$ to be sufficiently large compared to the tunnel coupling $\Delta_c$. This can be done by either careful engineering of the QDs to result in larger $\delta_o$ or by reducing $\Delta_c$ via the use of tunnel barrier gates. We note, the methodology described here can also be extended to other QD systems such as readout of spin-charge hybrid qubits \cite{kim2014,jang2021} where the two states of the computational basis experience distinct anticrossings separated from each other in detuning.


\section{Multivalued state measurements}\label{app:sec:qudit}

\begin{figure}
    \centering
    \includegraphics[width=0.95\linewidth]{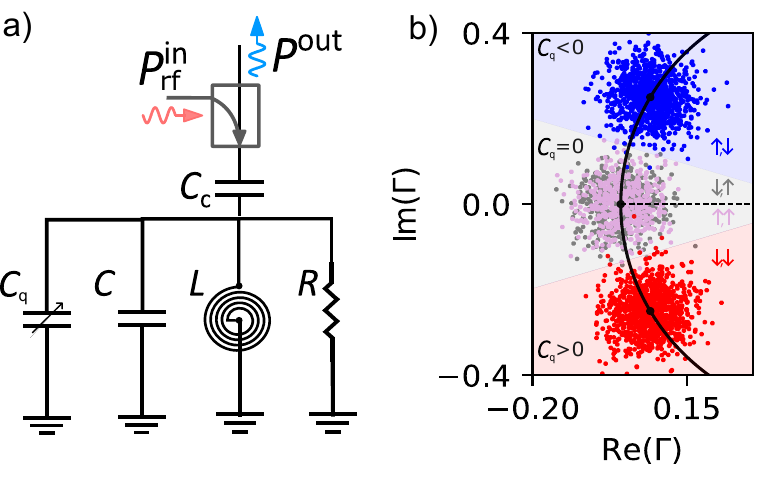}
    \caption{ a) Lumped element schematic of the resonator whose device capacitance, $\Delta C_\mathrm{q}$, depends on the spin state of the DQD. b) Impedance of a perfectly matched resonator as a function of frequency in the Smith chart (black circle). At the resonant frequency the resonator impedance ideally reaches the origin point (0,0). The three distinct clusters represent the simulated time-averaged impedance response of the resonator under three conditions: $\Delta C_\mathrm{q}=0$ (grey/pink), $\Delta C_\mathrm{q}>0$ (red) and $\Delta C_\mathrm{q}<0$ (blue). The three states can be classified by dividing the Smith chart into three regions, where each region corresponds to a different spin state.}
    \label{fig:appendix_qudit}
\end{figure}

As shown in Fig.~\ref{fig:readout-concept}i from the main manuscript, at detuning  $\epsilon=0.1$~meV  the quantum capacitance, $C_\text{q}$, takes on different values based on the spin states of the double quantum dot:

\begin{itemize}
    \item The state $\ket{\downarrow, \downarrow}$ has a negative quantum capacitance.
    \item The state $\ket{\downarrow, \uparrow}$  has a positive quantum capacitance.
    \item The states $\ket{\uparrow, \downarrow}$ and $\ket{\uparrow, \uparrow}$) both result in $C_\text{q}=0$.
\end{itemize}

To readout the spin state, the device is integrated into a resonator (\ref{fig:appendix_qudit}a) whose impedance response shifts with the value of $C_\text{q}$. \ref{fig:appendix_qudit}b shows simulated readout data for the three possible $C_\text{q}$ values. Each point represents a time-averaged readout trace, and due to noise, points cluster into distinct “lollipop-like” shapes corresponding to the different  $C_\text{q}$ values. This behaviour is akin to how multi-level qubits in superconducting systems respond~\cite{chen2023,champion2024}.

We can classify the resonator response in one of the three states by dividing the signal in the IQ plane into three regions separated by threshold lines. This method is similar to the thresholding approach used in Pauli spin blockade but extends into two dimensions and includes two thresholds to account for the three spin states. Beyond thresholding, more advanced machine learning techniques, such as support vector machines or neural networks, could enhance classification~\cite{chen2023,wang2024systematic}.

For a reliable assessment of classification performance in this \textit{tri-state} setup, the standard concept of state fidelity must be slightly modified. Borrowing from machine learning, the magnitude \textit{recall} is introduced to measure the classification accuracy for each state independently: recall for a specific state is defined as 1 - (misclassified traces for that state / total traces for that state). Additionally, the visibility of the readout remains defined as in the classical two-state scenario, capturing overall readout accuracy as 1 - (total errors / total measurements).


\section{T$^-$/S anticrossing detuning dependence with Magnetic field} \label{app:sec:Readout-epi-vs-B}

\begin{figure}
    \centering
    \includegraphics[width=\linewidth]{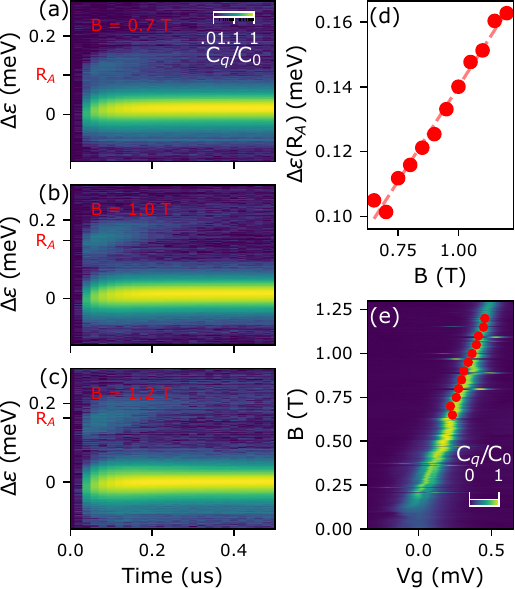}
    \caption{Characterisation of Readout point location for spin-orbit lifted readout against magnetic field: (a-c) Capacitive signal vs readout time like that in Fig. \ref{fig:readout-proof-of-principle}d at three different fields. Note that the short-lived excited state signal (marked with R$_A$) shifts with respect to the long-lived ground state signal. (d) Center of excited state signal location relative to the ground state signal as a function magnetic field. The dashed line has a slope of $\partial\varepsilon/\partial B = \overline{g}\mu_B$ with $\overline{g}$ of 2.01~$\pm$~.08. (e) The line in d overlayed on the magneto-spectroscopy data shown in Fig. \ref{fig:readout-concept}d showing a good match.}
    \label{fig:app:Readout-epi-vs-B}
\end{figure}

To further characterise the behaviour of the readout in the T$^-$/S coupled regime of ICT A, we characterise the location of the excited ($\uparrow_B,\downarrow_D$)/S anticrossing as a function of magnetic field. We carry out readout experiments like those described in section \ref{sec:ST-readout-demo} and measure the separation of the ground state signal arising from the T$^-$/S anticrossing from the excited state signal arising from the ($\uparrow_B,\downarrow_D$)/S anticrossing (Fig. \ref{fig:app:Readout-epi-vs-B}a-c).
We find this dependency to be approximately linear, with a slope given by $\Delta \varepsilon = \hat{g}\mu_B B$ from which we extract $\overline{g}$ of 2.01~$\pm$~.08, in line with what was extracted from magneto-spectroscopy (Fig. \ref{fig:app:Readout-epi-vs-B}d,e).

\section{Improvements required for single shot readout beyond the fault tolerance threshold}\label{app:sec:fidelity}

\begin{figure}[htb!]
    \centering
    \includegraphics[width=\linewidth]{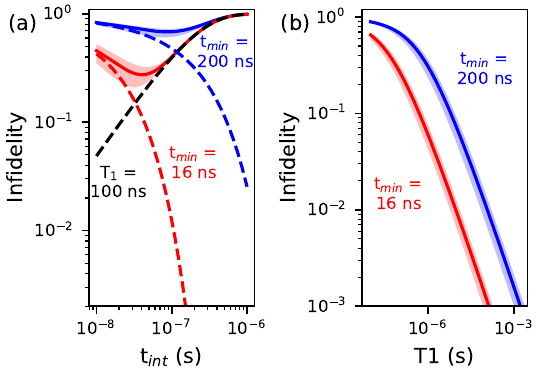}
    \caption{Readout fidelity as a function of $t_\text{int}$, $t_\text{min}$ and $T_1$ as calculated by Eq.~\ref{eq:fidelity-from-tmin-t1}. (a) Infidelity against $t_\text{int}$ calculated using the experimental values of $t_\text{min}$ and $T_1$. The plot shows optima where errors from $T_1$ and insufficient SNR balance. (b) Optimum Fidelity achievable as a function of $T_1$ for $t_\text{min}$ = 200~ns (blue) and 16~ns (red). The shaded areas represents the uncertainty the infidelity due to the uncertainty in the measured $t_\text{min}$.}
    \label{fig:fid-vs-T1-tmin}
\end{figure}

For practical quantum computing, qubit readout with a fidelity exceeding the error-correction threshold, e.g. about 99\% for the surface code \cite{raussendorf2007}, must be achieved. To estimate the requirements for such fidelity, we shall inspect the equation for the readout fidelity assuming simple box-car integration \cite{gambetta2007, Gunyho2024}:

\begin{equation}\label{eq:fidelity-from-tmin-t1}
    F = \text{erf}\left[\sqrt{\frac{t_\text{int}}{2t_\text{min}}}  \right] e^{\frac{-t_\text{int}}{2T_1}},
\end{equation}

\noindent where $\text{erf}(x)$ is the error function, $t_\text{int}$ is the integration time, and $t_\text{min}$ is the minimum integration time which is a measure of the electrical sensitivity of the measurement apparatus, defined as the integration time where the signal-to-noise ratio (SNR) is unity. We use here the fidelity as defined by Ref.~\cite{gambetta2007} where the fidelity is defined between 0 and 1. Note that this equation does not take into account the finite bandwidth of the readout resonator, which could lead to a lower fidelity at low $t_\text{int}$ due to the resonator ring-up.

Equation~\ref{eq:fidelity-from-tmin-t1} describes two opposing factors that impact fidelity: increasing $t_\text{int}$ results in an improved electrical SNR with a rate determined by $t_\text{min}$, but also increases the probability of relaxation of the state occurring during the measurement time, with a rate determined by $T_1$. This leads to an optimum $t_\text{int}$ which can be found given a $t_\text{min}$ and $T_1$, see \ref{fig:fid-vs-T1-tmin}. To improve fidelity, one must therefore improve these two metrics.

First, we estimate $t_\text{min}$ for the measurements in Fig.~\ref{fig:readout-concept}~d (see Supplementary~\ref{app:tmin-estimation}) and find $t_\text{min}$ is 200~$\pm$~80~ns and 16~$\pm$~6~ns for the ($\uparrow_B,\downarrow_D$) - S$_{2,0}$ and the ($\downarrow_B,\downarrow_D$) - S$_{2,0}$ transitions, respectively. We attribute the large difference in $t_\text{min}$ to the lower tunnel coupling. Particularly, $\Delta_\text{sf} \sim$ 3.4~GHz approaches the resonator frequency of 2.1~Ghz which maximises the resonator response \cite{peri2024a,Mizuta2017}. While our system lacked control of the tunnel coupling, $\Delta_\text{sc}$ could be optimised to improve the readout time. Further improvements in $t_\text{min}$ may be achieved by improved resonator design, for example by increasing the frequency, internal quality factor or the fractional change in capacitance \cite{Vigneau2023}. The use of quantum-limited amplifiers such as Josephson parametric amplifiers (JPAs) or kinetic inductance parametric amplifiers (KIPAs) could further reduce the noise temperature compared to that of the cryogenic high-electron-mobility transistor (HEMT) amplifier used in this experiment ($T_{n,\text{HEMT}} \sim$ 2.5~K) by up to an order of magnitude \cite{Oakes2023}. To estimate the effect such improvements may have, we compare our measurements to the state-of-the-art in dispersive charge sensing performed using a high-frequency resonator and a JPA \cite{Stehlik2015}. Adjusting for the impact of the lever arm on readout signal using the load-aware metric $t_\text{min}\alpha^2$ \cite{vonhorstig2023}, we arrive at a representative $t_\text{min}$ of 2.4~ns.

Next, we discuss the impact of $T_1$. In the present set-up, readout fidelity is limited by the state decay due to a short $T_1$ at the readout point on the order of 100 ns, preventing single-shot readout. We hypothesize that the short $T_1$ at the readout point is caused by the strong spin-orbit coupling in our sample that enables charge noise to couple into the spin degrees of freedom~\cite{Burkard2023_Semispinsreview}. This is supported by the exponential dependence of $T_1$ with detuning away from the anticrossing (Fig.~\ref{fig:T1-vs-detuning}), where the spin-carrying charges have the largest dipole and therefore are most impacted by charge noise. Further, although a detailed study of this was not carried out, the $T_1$ at the readout point shows no significant dependence with magnetic field -- see data presented in Supplementary \ref{app:sec:Readout-epi-vs-B}. This indicates that $T_1$ is likely not limited by phonon-mediated decay (1/T$_1$ $\propto B^7$) or due to fixed external magnetic field gradient or hyperfine-induced relaxation (both 1/T$_1$ $\propto B^5$) \cite{Burkard2023_Semispinsreview}. To reduce charge noise, improvements in the fabrication of the nanowire transistor could be of help, such as replacing the high-$\kappa$ gate dielectric, which is known to produce excessive charge noise \cite{Ibberson2018}, with SiO$_2$.

In \ref{fig:fid-vs-T1-tmin}b, we plot the measurement infidelity ($1-F$) at the optimum $t_\text{int}$ as a function $T_1$ for $t_\text{min}$ = 200~ns and 16~ns. Extracting the required $T_1$ to achieve a fidelity of 99\%, we arrive at 120~$\pm$~50~$\mu$s, 9.5~$\pm~$3~$\mu$s and, for the extrapolation of $t_\text{min}$ = 2.4~ns from Ref.~\cite{Stehlik2015}, 1.4~$\mu$s. While these $T_1$ are significantly longer than what was measured in this system, they are on the order of or lower than $T_1$ measured in spin systems measured using direct dispersive methods \cite{Zheng2019} as well as hole spin systems \cite{Hendrickx2021,Piot2022,Crippa2019}.

\section{Estimation of minimum measurement time.}\label{app:tmin-estimation}
We estimate $t_\text{min}$ from the data in the magneto-spectroscopy in \ref{fig:readout-concept} using the SNR of the resonator response. Extrapolating using a white noise model, this is given by \cite{Vigneau2023}

\begin{equation}
    t_\text{min} = t_{NE}/\text{SNR}^2,
\end{equation}
\noindent where the SNR is calculated from the peak amplitude over the noise, taken as the standard deviation of the signal at far detuning, and $t_{NE}$ is the noise equivalent integration time given by

\begin{equation}
    t_{NE} = n_\text{avg}/2\Delta f,
\end{equation}

\noindent which takes into account the number of averages ($n_\text{avg}$ = 200 in the magneto-spectroscopy dataset), as well as the equivalent noise bandwidth ($\Delta f = 1.57 f_c$ for a single-node low-pass filter with cut-of frequency $f_c$ = 30~kHz in this case), resulting in $t_{NE}$ of approximately 2.13~ms. Note that the long integration time is likely to introduce a reduction in the SNR due to charge noise \cite{vonhorstig2023}, meaning the true $t_\text{min}$ is likely to be lower.

Using this method for the data in Fig.~\ref{fig:readout-concept}~d, we estimate the $t_\text{min}$ for the response of the ($\uparrow_B,\downarrow_D$) - S$_{2,0}$ and the ($\downarrow_B,\downarrow_D$) - S$_{2,0}$ transitions by sampling the extrapolated $t_\text{min}$ at magnetic fields below 50~mT and at magnetic fields above 600~mT respectively, which ensures that the resonator response purely arises from the desired states. For these anticrossings, the resulting $t_\text{min}$ is 200~$\pm~$80~ns and 16~$\pm~$6~ns respectively. The large errors arise due to the extrapolation to low integration times from a single point.

\section{Impact of anticrossing overlap on dispersive and charge sensed readout visibility}\label{app:sec:readout-visibility-separation}

\begin{figure}[tb!]
    \centering
    \includegraphics[width=\linewidth]{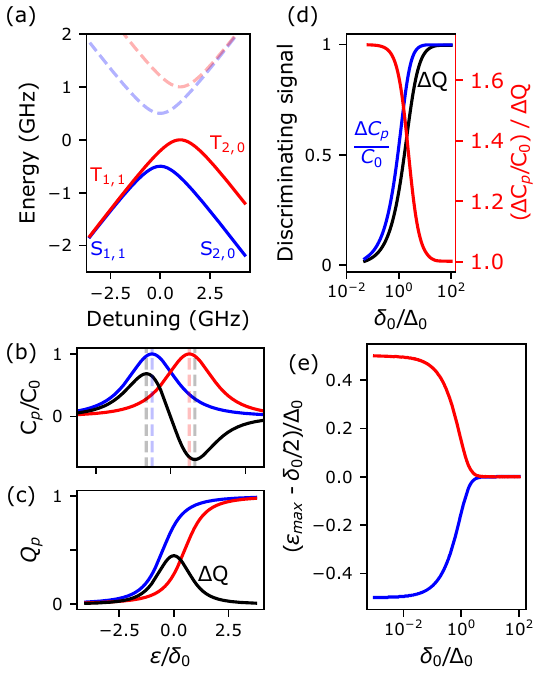}
    \caption{Comparison of impact of anticrossing overlap for charge sensed SB and dispersive sensing. (a) Energy-level diagram an orbitally lifted with $\Delta_\text{sc} = \delta_\text{o}$ = 1~GHz. (b, c) Capacitance $C_p$ (i.e. polarisability) and charge polarisation $Q_p$ plotted against detuning for the parameters plotted in (a). The S$_{1,1}$and S$_{0,2}$, and T$_{1,1}$/T$_{0,2}$ states are plotted in blue and red respectively. The black lines show the signal contrast of $\Delta C_p$ and $\Delta Q_p$. (d) Maximum $\Delta C_p$ and $\Delta Q$, as well as their normalised ratio plotted against orbital energy $\delta_\text{o}$. (e) Detuning-location of maximum $\Delta C_p$ plotted as offset from the bare anticrossing location, as a function of anticrossing separation $\delta_0$.}
    \label{fig:psb-comparison-orb}
\end{figure}

As discussed in Supplementary~\ref{app:sec:disc-convet-PSB-limits} and represented in \ref{fig:psb-comparison-orb}~b and c, an overlap in detuning of the relevant spin produces a loss of signal contrast for both charge-sensed and dispersive readout. This overlap impacts, for example, the signal contrast in the experiment in Section~\ref{sec:Excited_State}. In this Supplementary, we shall quantify the impact of signal loss as a function of anticrossing separation for both dispersive readout and conventional charge-sensed SB, see Fig~\ref{fig:psb-comparison-orb}d-e.

For the sake of simplicity, we examine the case of an orbitally lifted SB at high magnetic field (so the ($\downarrow,\downarrow$)/$S_{2,0}$ anticrossing can be ignored), where the orbital states are separated by energy $\delta_0$ and we have $\Delta_\text{sf}$ = 0 and $\Delta_\text{sc}$ = $\Delta_\text{0}$. This produces two anticrossings for the $\downarrow,\downarrow$ and $\uparrow,\downarrow$ states with equal coupling to their respective (0,2) states, separated by $\delta_0$, which we can then adjust to evaluate the impact of the separation of the anticrossings. Note that a similar approach may be taken to evaluate the impact in the spin-orbit lifted SB case.

In \ref{fig:psb-comparison-orb}b-c, we plot the normalized capacitance $C_p/C_0$ (b) and the charge polarization $Q_p$ (c) for the $\uparrow,\downarrow$ and $\downarrow,\downarrow$ states (blue and red solid lines) as a function of reduced detuning for the case $\delta_0$ = $\Delta_\text{0}$. Furthermore, we plot the signal contrast $\Delta C_p$ and $\Delta Q_p$ (solid black lines). Note that in the upper panel for $C_p$, both the maximum signal contrast amplitude ($|\Delta C_p|/C_0$), and the detuning location at which it occurs ($\varepsilon_{max}$, vertical black dashed lines) differ from that of the pure signal (vertical blue and red dashed lines). 

In \ref{fig:psb-comparison-orb}d, we vary $\delta_0$ and plot the maximum signal contrast of $|\Delta C_p|/C_0$ (blue) and $\Delta Q_p$ (black). We also plot the ratio of the signal contrasts for the two cases (red). In either case, a signal reduction occurs when $\delta_0$ is of the order of $\Delta_0$. However, while both cases are affected, $|\Delta C_p|/C_0$ is more resilient, as shown by the red line. This difference originates from the fact that, unlike the dispersive case, the optimum readout location for the charge-sensed case is located at the midpoint of the anticrossings, increasing the impact of the overlap. 

For the dispersive case, the location at which the maximum $\Delta C_p$ occurs is a function of $\delta_0$ as shown in \ref{fig:psb-comparison-orb}e. When $\delta_0$ becomes of the order of $\Delta_0$, the $\varepsilon_{max}$ for the two states are pushed outward, plateauing at a value of $\delta_0$ for $\delta_0$/$\Delta_0$ $\ll$ 1 (ie $\varepsilon_{max}$ becomes entirely determined by $\delta_0$). These results highlight the importance of calibrating the readout points for low-energy anticrossing separations.

To improve $\delta_0/\Delta_0$, two general approaches can be taken: (i) in the spin-orbit case, the magnetic field can be adjusted to increase the separation $\delta_0$ of the $\uparrow,\uparrow$ and $\downarrow,\uparrow$ anticrossings with S$_{2,0}$. (ii) Tunnel barriers (if present) can be used to decrease tunnel coupling $\Delta_0$ which increases the ratio of $\delta_0/\Delta_0$. With the latter approach, care should be taken not to reduce the $\Delta$ of the state to be measured below the resonator frequency $f_r$ when utilizing gate dispersive readout, as this reduces the signal amplitude~\cite{peri2024a}. A detailed description of the various readout points and their separations for the two systems described in this work can be found in Supplementary~\ref{app:sec:readout-locations}.

\bibliography{General_clean}
